\documentclass{article}

\usepackage{amsmath}  
\usepackage{amssymb}  
\usepackage{feynmf}   
\usepackage{texdraw}  

\begin{document}



\title{Effective Field Theory Dimensional Regularization}
\author{Dirk Lehmann\thanks{\texttt{dlehmann@jlab.org}}, 
  Gary Pr\'{e}zeau\thanks{\texttt{prezeau@jlab.org}}\\
  \textit{Department~of~Physics, Hampton~University,}\\ 
  \textit{Hampton, VA~23668, USA}\thanks{Mail address: Thomas Jefferson National Accelerator
    Facility, 12000 Jefferson Avenue, M.S.~12H2, Newport News, VA
    23606, USA}}

\date{\today}

\newcommand{\I}{\text{i}} 
\newcommand{\D}{\text{d}} 
\renewcommand{\bar}{\overline} 


\begin{fmffile}{graphs} 

  \def\SelfenergyIntro{%
    \parbox[][70pt][t]{120pt}{%
      \begin{fmfgraph*}(120,80)
        \fmfpen{5} \fmfstraight
        \fmfleft{left_down,left,left_up} \fmfright{right_down,right,right_up}
        \fmf{fermion,tension=2}{left,m1} 
        \fmf{fermion,label=$M$,l.side=right}{m1,m2} 
        \fmf{fermion,tension=2}{m2,right}
        \fmffreeze
        \fmf{dashes,left=1,label=$m$,l.side=left}{m1,m2}
        \fmfdot{m1,m2} 
      \end{fmfgraph*}
      }
    }
  \def\TriangleGraph{%
    \parbox[][75pt][c]{120pt}{%
      \begin{fmfgraph*}(100,70)
        \fmfpen{5} \fmfstraight
        \fmfleft{lower_left,middle_left,upper_left} \fmfright{lower_right,middle_right,upper_right}
        \fmf{phantom}{upper_left,vertex_left} \fmf{phantom}{lower_left,vertex_left}
        \fmf{phantom}{vertex_left,upper_right} \fmf{phantom}{vertex_left,lower_right}
        \fmffreeze
        \fmf{phantom,tension=2}{upper_left,incoming_upper}
        \fmf{phantom,tension=2}{lower_left,incoming_lower} 
        \fmf{fermion}{incoming_upper,vertex_left} \fmf{fermion}{incoming_lower,vertex_left}
        \fmf{fermion,label=$P_{1}+k$,l.side=left}{vertex_left,vertex_upper}
        \fmf{fermion,label=$k-P_{2}$,l.side=left}{vertex_lower,vertex_left}  
        \fmf{fermion,tension=2}{vertex_upper,upper_right}
        \fmf{fermion,tension=2}{vertex_lower,lower_right} 
        \fmffreeze
        \fmf{scalar,label=$k$,l.side=left}{vertex_upper,vertex_lower}
        \fmf{phantom,label=$\scriptstyle 3$,l.side=right,l.dist=10}{vertex_upper,vertex_lower}
        \fmf{phantom,label=$\scriptstyle 1$,l.side=right,l.dist=10}{vertex_left,vertex_upper}
        \fmf{phantom,label=$\scriptstyle 2$,l.side=left,l.dist=10}{vertex_left,vertex_lower}  
        \fmfdot{vertex_left,vertex_upper,vertex_lower}
        \fmfv{label=$P_{1}$,l.angle=0}{upper_right} \fmfv{label=$P_{2}$,l.angle=0}{lower_right}
      \end{fmfgraph*}
      }
    }
  

  \def\BubbleGraph{%
    \parbox[][75pt][c]{120pt}{%
      \begin{fmfgraph*}(100,50)
        \fmfpen{5} \fmfstraight
        \fmfleft{left_down,left,left_up} \fmfright{right_down,right,right_up}
        \fmf{phantom,tension=4}{left,left_vertex}
        \fmf{plain,left=.5,label=$\scriptstyle 1$}{left_vertex,right_vertex}
        \fmf{plain,right=.5,label=$\scriptstyle 2$}{left_vertex,right_vertex}
        \fmf{phantom,tension=4}{right_vertex,right}
        \fmfdot{left_vertex,right_vertex} \fmffreeze
        \fmf{fermion}{left_down,left_vertex} \fmf{fermion}{left_up,left_vertex}
        \fmf{fermion}{right_vertex,right_down} \fmf{fermion}{right_vertex,right_up}
        \fmfv{label=$P_{1}$,l.angle=0}{right_up} \fmfv{label=$P_{2}$,l.angle=0}{right_down}
      \end{fmfgraph*}
      }
    }
  

  \def\SelfenergyGraphA{%
    \parbox[][75pt][c]{120pt}{%
      \begin{fmfgraph*}(100,50)
        \fmfpen{5} \fmfstraight
        \fmfleft{left_down,left,left_up} \fmfright{right_down,right,right_up}
        \fmf{fermion,tension=2}{left,m1} \fmf{plain}{m1,m2} \fmf{fermion,tension=2}{m2,right}
        \fmf{phantom}{left_down,d1} \fmf{phantom}{d1,right_down} \fmffreeze
        \fmf{dashes,left=1,label=$\scriptstyle 3$}{m1,m2}
        \fmf{phantom,label=$\scriptstyle 2$,l.side=right}{m1,m2}
        \fmfdot{m1,m2} \fmf{fermion}{left_down,m1} \fmf{fermion}{m1,d1}
        \fmfv{label=$P_{2}$,l.angle=0}{right} \fmfv{label=$P_{1}$,l.angle=0}{d1}
      \end{fmfgraph*}
      }
    }
  
 
  \def\SelfenergyGraphB{%
    \parbox[][75pt][c]{120pt}{%
      \begin{fmfgraph*}(100,50)
        \fmfpen{5} \fmfstraight
        \fmfleft{left_down,left,left_up} \fmfright{right_down,right,right_up}
        \fmf{fermion,tension=2}{left,m1} \fmf{plain}{m1,m2} \fmf{fermion,tension=2}{m2,right}
        \fmf{phantom}{left_down,d1} \fmf{phantom}{d1,right_down} \fmffreeze
        \fmf{dashes,left=1,label=$\scriptstyle 3$}{m1,m2}
        \fmf{phantom,label=$\scriptstyle 1$,l.side=right}{m1,m2}
        \fmfdot{m1,m2} \fmf{fermion}{left_down,m1} \fmf{fermion}{m1,d1}
        \fmfv{label=$P_{1}$,l.angle=0}{right} \fmfv{label=$P_{2}$,l.angle=0}{d1}
      \end{fmfgraph*} 
      }
    }
  
  
  \def\HeavyTadpoleGraph#1{%
    \parbox[][75pt][b]{55pt}{%
      \begin{fmfgraph*}(40,40)
        \fmfpen{5} \fmfstraight
        \fmfleft{left_down,left,left_up} \fmfright{right_down,right,right_up}
        \fmf{fermion}{left_down,vertex} \fmf{fermion}{left_up,vertex}
        \fmf{fermion}{vertex,right_down} \fmf{fermion}{vertex,right_up}
        \fmf{plain,right=90,tension=0.5,label=$\scriptstyle #1$,l.dist=10}{vertex,vertex}
        \fmfdot{vertex} 
        \fmfv{label=$P_{1}$,l.angle=0}{right_up} \fmfv{label=$P_{2}$,l.angle=0}{right_down}
      \end{fmfgraph*}  \vspace{8pt}
      }
    }
  
  
  \def\LightTadpoleGraph{%
    \parbox[][75pt][b]{55pt}{%
      \begin{fmfgraph*}(40,40)
        \fmfpen{5} \fmfstraight
        \fmfleft{left_down,left,left_up} \fmfright{right_down,right,right_up}
        \fmf{fermion}{left_down,vertex} \fmf{fermion}{left_up,vertex}
        \fmf{fermion}{vertex,right_down} \fmf{fermion}{vertex,right_up}
        \fmf{dashes,right=90,tension=0.5,label=$\scriptstyle 3$,l.dist=10}{vertex,vertex}
        \fmfdot{vertex}
        \fmfv{label=$P_{1}$,l.angle=0}{right_up} \fmfv{label=$P_{2}$,l.angle=0}{right_down}
      \end{fmfgraph*} \vspace{8pt}
      }
    }
  
  
  \def\TwoLoopGraph{%
    \parbox[][90pt][c]{150pt}{%
      \begin{center}
        \begin{fmfgraph}(120,80)
          \fmfpen{5} \fmfstraight
          \fmfleft{lower_left,middle_left,upper_left} \fmfright{lower_right,middle_right,upper_right}
          \fmf{phantom}{upper_left,vertex_left} \fmf{phantom}{lower_left,vertex_left}
          \fmf{phantom}{vertex_left,upper_right} \fmf{phantom}{vertex_left,lower_right}
          \fmffreeze
          \fmf{phantom,tension=2}{upper_left,incoming_upper} 
          \fmf{phantom,tension=2}{lower_left,incoming_lower}
          \fmf{fermion}{incoming_upper,vertex_left}
          \fmf{fermion}{incoming_lower,vertex_left}
          \fmf{plain}{vertex_left,vertex_upper}
          \fmf{plain}{vertex_left,vertex_lower}  
          \fmf{fermion,tension=2}{vertex_upper,upper_right}
          \fmf{fermion,tension=2}{vertex_lower,lower_right}
          \fmffreeze
          \fmf{dashes}{vertex_upper,vertex_lower}
          \fmf{plain,left=180,tension=0.6}{vertex_left,vertex_left}
          \fmfdot{vertex_left,vertex_upper,vertex_lower}
        \end{fmfgraph}
      \end{center}
      }
    }
  

  \def\RenormTriangleGraph{%
    \parbox[][90pt][c]{150pt}{%
      \begin{fmfgraph}(120,80)
        \fmfpen{5} \fmfstraight
        \fmfleft{lower_left,middle_left,upper_left} \fmfright{lower_right,middle_right,upper_right}
        \fmf{phantom}{upper_left,vertex_left} \fmf{phantom}{lower_left,vertex_left}
        \fmf{phantom}{vertex_left,upper_right} \fmf{phantom}{vertex_left,lower_right}
        \fmffreeze
        \fmf{phantom,tension=2}{upper_left,incoming_upper}
        \fmf{phantom,tension=2}{lower_left,incoming_lower}
        \fmf{fermion}{incoming_upper,vertex_left} \fmf{fermion}{incoming_lower,vertex_left}
        \fmf{plain}{vertex_left,vertex_upper}
        \fmf{plain}{vertex_left,vertex_lower}  
        \fmf{fermion,tension=2}{vertex_upper,upper_right} \fmf{fermion,tension=2}{vertex_lower,lower_right}
        \fmffreeze
        \fmf{dashes}{vertex_upper,vertex_lower}
        \fmfdot{vertex_upper,vertex_lower} \fmfv{deco.shape=square,deco.filled=gray50}{vertex_left}
      \end{fmfgraph}
      }
    }
  
  
  \def\GenSelfenergyGraph{%
    \parbox[][110pt][t]{150pt}{%
      \begin{fmfgraph*}(150,150)
        \fmfpen{5} \fmfcurved \fmfsurroundn{e}{8}
        \begin{fmffor}{n}{1}{1}{8} \fmf{phantom}{e[n],i[n]} \end{fmffor}
        \fmfcyclen{phantom,tension=1,right=1.2/7}{i}{8} \fmffreeze
        \fmf{fermion}{e5,i5}\fmf{plain}{i5,f1,f2,i1}\fmf{fermion}{i1,e1}
        \fmfn{dashes,right=1.2/7}{i}{5}
        \fmf{phantom}{e4,ee4}\fmf{dbl_dots,left=1.2/7}{ee4,ee2}\fmf{phantom}{ee2,e2}
        \fmf{photon}{e2,i2}\fmf{photon}{e4,i4} \fmf{phantom}{e6,h1,h2,e8}
        \fmffreeze \fmf{phantom}{h1,k1}\fmf{photon,tension=2}{k1,g1,f1}
        \fmf{phantom}{h2,k2}\fmf{photon,tension=2}{k2,g2,f2} \fmffreeze
        \fmf{phantom}{g1,gg1}\fmf{dbl_dots}{gg1,gg2}\fmf{phantom}{gg2,g2}
        \fmfdot{i1,i2,i4,i5}\fmfdotn{f}{2}
        \fmffreeze
        \fmf{phantom,label=$w_{j}$,l.side=left}{f1,f2}
        \fmf{phantom,right=1.2/7,label=$z_{i}$,l.side=left}{i[3],i[4]}
      \end{fmfgraph*}
      }
    }
  
  
  \def\SoftHeavyRingGraph#1{%
    \parbox[][110pt][b]{100pt}{%
      \begin{fmfgraph*}(100,100)
        \fmfpen{5} \fmfcurved \fmfsurroundn{e}{#1}
        \begin{fmffor}{n}{1}{1}{#1} \fmf{photon}{e[n],i[n]} \end{fmffor}
        \fmfcyclen{plain,tension=#1/8,right=1.2/(#1-1)}{i}{#1}
        \fmfdotn{i}{#1}
        \fmfv{label=$q_{i}$,l.angle=-20,l.dist=40}{i[3]}
        \end{fmfgraph*}
      }
    }

  
  \def\GenOneLoopGraphA#1{%
    \parbox[][210pt][c]{200pt}{%
      \begin{fmfgraph}(200,200)
        \fmfpen{5} \fmfcurved \fmfsurroundn{e}{6*#1}
        \begin{fmffor}{n}{1}{3}{6*#1}\fmf{plain,width=15}{e[n],m[n],i[n]}\end{fmffor}
        \begin{fmffor}{n}{2}{3}{6*#1}
          \fmf{phantom}{e[n],p[n]} \fmf{photon}{p[n],m[n],i[n]}
        \end{fmffor}
        \begin{fmffor}{n}{3}{3}{6*#1}
          \fmf{phantom}{e[n],p[n]} \fmf{photon}{p[n],m[n],i[n]}
        \end{fmffor}
        \fmfcyclen{dashes,tension=#1*6/8,right=1.2/(6*#1-1)}{i}{6*#1}
        \fmffreeze
        \begin{fmffor}{n}{1}{6}{6*#1}
          \fmf{plain,tension=#1*6/8,right=1.2/(6*#1-1),width=15}{i[n],i[n+1],i[n+2],i[n+3]}
        \end{fmffor}
        \fmffreeze
        \begin{fmffor}{n}{2}{3}{6*#1}
          \fmf{phantom}{m[n],mm[n]}
          \fmf{dbl_dots,tension=#1*6/8,right=1.2/(6*#1-1)}{mm[n],mm[n+1]}
          \fmf{phantom}{mm[n+1],m[n+1]}
        \end{fmffor}
        \fmfdotn{i}{6*#1}
      \end{fmfgraph}
      }
    }


  \def\RegularOneHeavyGraph{%
    \parbox[][110pt][t]{100pt}{%
      \begin{fmfgraph}(100,100)
        \fmfpen{5} \fmfcurved \fmfsurroundn{e}{8}
        \begin{fmffor}{n}{1}{1}{8} \fmf{phantom,tension=0.6}{e[n],i[n]} \end{fmffor}
        \fmfcyclen{plain,tension=1,right=1.2/7}{i}{8} \fmffreeze
        \begin{fmfsubgraph}(0,0)(w,0.5h)
          \fmfpen{5}\fmfleft{a}\fmfright{c}\fmf{fermion}{a,b,c}
        \end{fmfsubgraph}\fmffreeze
        \fmf{photon,tension=3}{e5,i5}\fmf{photon}{e1,i1}
        \fmf{photon,tension=3}{e4,m4}\fmf{photon}{m4,i4}
        \fmf{photon,tension=3}{e2,m2}\fmf{photon}{m2,i2}\fmffreeze
        \fmf{phantom}{m4,mm4}\fmf{dbl_dots,left=0.2}{mm4,mm2}\fmf{phantom}{mm2,m2}
        \fmffreeze
        \fmf{photon}{e6,m7,i7}\fmf{photon}{e8,m8,i7}\fmffreeze
        \fmf{phantom}{m7,mm7}\fmf{dbl_dots,right=0.3}{mm7,mm8}\fmf{phantom}{mm8,m8}
       \fmfdot{i1,i2,i4,i5,i7}
      \end{fmfgraph}
      }
    }  

  
  \def\GenBoxGraph{%
    \parbox[][110pt][c]{150pt}{%
      \begin{fmfgraph*}(150,100)
        \fmfpen{5} \fmfstraight
        \fmfleftn{l}{6} \fmfrightn{r}{6} \fmftopn{o}{6}  \fmfbottomn{u}{6}
        \fmf{plain}{l5,a1,a2,a3,a4,r5} \fmf{plain}{l2,b1,b2,b3,b4,r2}
        \fmffreeze
        \fmf{phantom_arrow}{l5,a1}\fmf{phantom_arrow}{a4,r5}
        \fmf{phantom_arrow}{l2,b1}\fmf{phantom_arrow}{b4,r2}
        \fmf{dashes}{a1,c1,c2,b1}\fmf{dashes}{a4,d1,d2,b4}
        \fmffreeze
        \fmf{photon}{l4,c3,c1}\fmf{photon}{l3,c4,c2}
        \fmf{photon}{d1,d3,r4}\fmf{photon}{d2,d4,r3}
        \fmf{photon}{o3,f1,a2}\fmf{photon}{o4,f2,a3}
        \fmf{photon}{u3,e1,b2}\fmf{photon}{u4,e2,b3}
        \fmffreeze
        \fmf{phantom}{f1,ff1}\fmf{dbl_dots}{ff1,ff2}\fmf{phantom}{ff2,f2}
        \fmf{phantom}{e1,ee1}\fmf{dbl_dots}{ee1,ee2}\fmf{phantom}{ee2,e2}
        \fmf{phantom}{c3,cc3}\fmf{dbl_dots}{cc3,cc4}\fmf{phantom}{cc4,c4}
        \fmf{phantom}{d3,dd3}\fmf{dbl_dots}{dd3,dd4}\fmf{phantom}{dd4,d4}
        \fmfdotn{a}{4}\fmfdotn{b}{4}\fmfdot{c1,c2,d1,d2}
        \fmffreeze
        \fmf{phantom,label=$x_{i}$,l.side=right}{a2,a3}
        \fmf{phantom,label=$y_{j}$,l.side=left}{b2,b3}
        \fmf{phantom,label=$z_{k}$,l.side=left}{c1,c2}
      \end{fmfgraph*}
      }
    }
  
  
  \def\GenCrossBoxGraph{%
    \parbox[][110pt][c]{150pt}{%
      \begin{fmfgraph*}(150,100)
        \fmfpen{5} \fmfstraight
        \fmfleftn{l}{6} \fmfrightn{r}{6} \fmftopn{o}{6}  \fmfbottomn{u}{6}
        \fmf{plain}{l5,a1,a2,a3,a4,r5} \fmf{plain}{l2,b1,b2,b3,b4,r2}
        \fmffreeze 
        \fmf{phantom_arrow}{l5,a1}\fmf{phantom_arrow}{a4,r5}
        \fmf{phantom_arrow}{l2,b1}\fmf{phantom_arrow}{b4,r2}
        \fmffreeze 
        \fmf{photon}{o3,f1,a2} \fmf{photon}{o4,f2,a3}
        \fmf{photon}{u3,e1,b2} \fmf{photon}{u4,e2,b3}
        \fmffreeze 
        \fmf{phantom}{f1,ff1}\fmf{dbl_dots}{ff1,ff2}\fmf{phantom}{ff2,f2}
        \fmf{phantom}{e1,ee1}\fmf{dbl_dots}{ee1,ee2}\fmf{phantom}{ee2,e2}
        \fmffreeze 
        \fmf{dashes,tension=3}{a1,c1}\fmf{dashes,tension=3}{c1,c2}\fmf{dashes}{c2,b4}
        \fmf{phantom,tension=3}{l5,cc1}\fmf{phantom,tension=3}{cc1,cc2}\fmf{phantom}{cc2,b3}
        \fmf{dashes}{b1,d1}\fmf{dashes,tension=3}{d1,d2}\fmf{dashes,tension=3}{d2,a4}
        \fmf{phantom}{b2,dd1}\fmf{phantom,tension=3}{dd1,dd2}\fmf{phantom,tension=3}{dd2,r5}
        \fmffreeze 
        \fmf{photon}{cc1,ccc1,c1}\fmf{photon}{cc2,ccc2,c2}
        \fmf{photon}{dd1,ddd1,d1}\fmf{photon}{dd2,ddd2,d2}
        \fmffreeze 
        \fmf{phantom}{ccc1,C}\fmf{dbl_dots}{C,CC}\fmf{phantom}{CC,ccc2}
        \fmf{phantom}{ddd1,D}\fmf{dbl_dots}{D,DD}\fmf{phantom}{DD,ddd2}
        \fmfdotn{a}{4}\fmfdotn{b}{4}\fmfdot{c1,c2,d1,d2} 
      \end{fmfgraph*}
      }
    }

  \def\BoxGraph{%
    \parbox[][60pt][c]{120pt}{%
      \begin{fmfgraph*}(120,50)
        \fmfpen{5} \fmfstraight
        \fmfleft{lower_left,upper_left} \fmfright{lower_right,upper_right}
        \fmf{fermion,tension=3}{upper_left,vertex_upper_left}
        \fmf{fermion,tension=3}{lower_left,vertex_lower_left}
        \fmf{plain,label=$\scriptstyle 1$,l.side=left,l.dist=10}{vertex_upper_left,vertex_upper_right}
        \fmf{plain,label=$\scriptstyle 2$,l.side=left,l.dist=10}{vertex_lower_left,vertex_lower_right}
        \fmf{fermion,tension=3}{vertex_upper_right,upper_right}
        \fmf{fermion,tension=3}{vertex_lower_right,lower_right}
        \fmffreeze \fmfpen{5}
        \fmf{dashes,width=5}{vertex_lower_left,vertex_upper_left}
        \fmf{dashes,width=5}{vertex_upper_right,vertex_lower_right}
        \fmfdot{vertex_upper_left,vertex_lower_left,vertex_upper_right,vertex_lower_right}
      \end{fmfgraph*}
      }
    }


  \def\CrossBoxGraph{%
    \parbox[][60pt][c]{120pt}{%
      \begin{fmfgraph*}(120,50)
        \fmfpen{5} \fmfstraight
        \fmfleft{lower_left,upper_left} \fmfright{lower_right,upper_right}
        \fmf{fermion,tension=3}{upper_left,vertex_upper_left}
        \fmf{fermion,tension=3}{lower_left,vertex_lower_left}
        \fmf{plain,label=$\scriptstyle 1$,l.side=left,l.dist=10}{vertex_upper_left,vertex_upper_right}
        \fmf{plain,label=$\scriptstyle 2$,l.side=left,l.dist=10}{vertex_lower_left,vertex_lower_right}
        \fmf{fermion,tension=3}{vertex_upper_right,upper_right}
        \fmf{fermion,tension=3}{vertex_lower_right,lower_right}
        \fmffreeze \fmfpen{5}
        \fmf{dashes,width=5,rubout}{vertex_lower_left,vertex_upper_right}
        \fmf{dashes,width=5}{vertex_upper_left,vertex_lower_right}
        \fmfdot{vertex_upper_left,vertex_lower_left,vertex_upper_right,vertex_lower_right}
      \end{fmfgraph*}
      }
    }

  
  \def\ThreeHeavyGraph{%
    \parbox[][60pt][c]{120pt}{%
      \begin{fmfgraph}(120,50)
        \fmfpen{5} \fmfstraight
        \fmfleft{c1,b1,a1}\fmfright{c5,b5,a5}
        \fmfn{phantom}{a}{5}\fmfn{phantom}{b}{5}\fmfn{phantom}{c}{5}
        \fmffreeze
        \fmf{dashes}{a2,b2}\fmf{dashes}{b3,c3}\fmf{dashes}{c4,a4}
        \fmf{fermion}{a1,a2}\fmf{plain}{a2,a4}\fmf{fermion}{a4,a5}
        \fmf{fermion}{b1,b2}\fmf{plain}{b2,b3,b4}\fmf{fermion}{b4,b5}
        \fmf{fermion}{c1,c2}\fmf{plain}{c2,c3,c4}\fmf{fermion}{c4,c5} 
        \fmfdot{a2,a4,b2,b3,c3,c4}
      \end{fmfgraph}
      }
    }

  
  \def\GenOneLoopGraphB#1{%
    \parbox[][210pt][c]{200pt}{%
      \begin{fmfgraph}(200,200)
        \fmfpen{5} \fmfcurved \fmfsurroundn{e}{4*#1}
        \begin{fmffor}{n}{1}{2}{4*#1}\fmf{plain,width=15}{e[n],m[n],i[n]}\end{fmffor}
        \begin{fmffor}{n}{2}{2}{4*#1}
          \fmf{phantom}{e[n],p[n]} \fmf{photon}{p[n],m[n],i[n]}
        \end{fmffor}
        \fmfcyclen{dashes,tension=4*#1/8,right=1.2/(4*#1-1)}{i}{4*#1}
        \fmffreeze
        \begin{fmffor}{n}{1}{4}{4*#1}
          \fmf{plain,tension=4*#1/8,right=1.2/(4*#1-1),width=15}{i[n],i[n+1],i[n+2]}
        \end{fmffor}
        \fmfdotn{i}{4*#1}
      \end{fmfgraph}
      }
    }


  \def\LandauLSG{%
    \linewd 0.2 \setgray 0.8
    \lpatt() \rmove(-0.3 -0.3) \rlvec(1.6 1.6) \rmove(-1.3 -1.3) 
    \rmove(-0.3 0.3) \rlvec(1.6 -1.6) \rmove(-0.3 0.3) 
    \lpatt(0.2 0.4) \rlvec(0 2) \rmove(-1 -1) \lpatt()
    \linewd 0.1 \setgray 0
    \lpatt() \rmove(-0.3 -0.3) \rlvec(1.6 1.6) \rmove(-1.3 -1.3) 
    \rmove(-0.3 0.3) \rlvec(1.6 -1.6) \rmove(-0.3 0.3) 
    \lpatt(0.2 0.4) \rlvec(0 2) \rmove(-1 -1) \lpatt()
    }  
  \def\LandauSEG{%
    \lpatt() \rmove(-0.3 0) \rlvec(2.6 0) \rmove(-1.3 0) \lpatt(0.2
    0.4) \larc r:1 sd:0 ed:180 \lpatt() 
    }
  \def\LandauBG{%
    \lpatt() \linewd 0.2 \setgray 1 
    \rmove(0 -0.8) \larc r:1.3 sd:0 ed:180 
    \rmove(0 1.6) \larc r:1.3 sd:180 ed:360 \linewd 0.1 \setgray 0 \rmove(0 -0.8)
    \lpatt() \rmove(0 -0.8) \larc r:1.3 sd:0 ed:180 
    \rmove(0 1.6) \larc r:1.3 sd:180 ed:360
    }
  \def\LandauHTG{%
    \lpatt() \fcir f:0 r:0.25 \rmove(0 1) \larc r:1 sd:0 ed:360 
    }
  \def\LandauLTG{%
    \lpatt(0.2 0.3) \fcir f:0 r:0.25 \rmove(0 1) \larc r:1 sd:0 ed:360 \lpatt() 
    }
  \def\LandauArrow#1{%
    \linewd 0.05 \lpatt() \arrowheadsize l:0.6 w:0.25 \arrowheadtype
    t:V \avec(#1) \lpatt() \linewd 0.1 
    }
  \def\LandauCoords{%
    \setunitscale 0.7
    \arrowheadtype t:F \arrowheadsize l:2 w:1
    \linewd 0.3 \rmove(0 8) \rlvec(-6 -14) \rlvec(12 0) \rlvec(-6 14)
    \lfill f:0.8 \rmove(0 -8) 
    \linewd 0.2 \lpatt(0.5 1) \rlvec(0 8) \lpatt()\ravec(0 7)
    \textref h:C v:B \htext{$\alpha_{3}$} \rmove(0 -15)  
    \linewd 0.2 \lpatt(0.5 1) \rlvec(-6 -6)\lpatt()\ravec(-4 -4)
    \textref h:R v:T \htext{$\alpha_{1}$} \rmove(10 10)  
    \linewd 0.2 \lpatt(0.5 1) \rlvec(6 -6) \lpatt() \ravec(4 -4)
    \textref h:L v:T \htext{$\alpha_{2}$} \rmove(-10 10) 
    \linewd 0.7 \rmove(-18 -16) \rlvec(35 0) \rlvec(0 35) \rlvec(-35
    0) \rlvec(0 -35) \rmove(18 16) 
    \setunitscale 3 \linewd 0.1
    }
  \def\LandauLeadA{%
    \move(13.51 0) \lvec(13.68 1) \lvec(13.85 2) \lvec(14.02 3) \lvec(14.20 4) 
    \lvec(14.37 5) \lvec(14.54 6) \lvec(14.71 7) \lvec(14.89 8) \lvec(15.06 9)
    \lvec(15.23 10) \lvec(15.40 11) \lvec(15.58 12) \lvec(15.75 13) \lvec(15.92 14) 
    \lvec(16.09 15) \lvec(16.27 16) \lvec(16.44 17) \lvec(16.61 18) \lvec(16.78 19) 
    \lvec(16.96 20) \lvec(17.13 21) \lvec(17.30 22) \lvec(17.47 23) \lvec(17.65 24) 
    \lvec(17.82 25) \lvec(17.99 26) 
    \lvec(19.35 26) \lvec(18.85 25) \lvec(18.37 24) \lvec(17.92 23) \lvec(17.50 22)
    \lvec(17.13 21) \lvec(16.80 20) \lvec(16.54 19) \lvec(16.35 18) \lvec(16.23 17)
    \lvec(16.17 16) \lvec(16.14 15) \lvec(16.11 14) \lvec(16.04 13) \lvec(15.92 12)
    \lvec(15.73 11) \lvec(15.47 10) \lvec(15.13 9) \lvec(14.72 8) \lvec(14.23 7) 
    \lvec(13.67 6) \lvec(13.06 5) \lvec(12.41 4) \lvec(11.72 3) \lvec(11.00 2) 
    \lvec(10.26 1) \lvec(9.51 0) \lvec(13.51 0) \ifill f:0.5
    }
  \def\LandauLeadB{%
    \move(16.49 0) \lvec(16.32 1) \lvec(16.15 2) \lvec(15.98 3) \lvec(15.80 4) 
    \lvec(15.63 5) \lvec(15.46 6) \lvec(15.29 7) \lvec(15.11 8) \lvec(14.94 9)
    \lvec(14.77 10) \lvec(14.60 11) \lvec(14.42 12) \lvec(14.25 13) \lvec(14.08 14) 
    \lvec(13.91 15) \lvec(13.73 16) \lvec(13.56 17) \lvec(13.39 18) \lvec(13.22 19) 
    \lvec(13.04 20) \lvec(12.87 21) \lvec(12.70 22) \lvec(12.53 23) \lvec(12.35 24) 
    \lvec(12.18 25) \lvec(12.01 26) 
    \lvec(10.65 26) \lvec(11.15 25) \lvec(11.63 24) \lvec(12.08 23) \lvec(12.50 22) 
    \lvec(12.87 21) \lvec(13.20 20) \lvec(13.46 19) \lvec(13.65 18) \lvec(13.77 17) 
    \lvec(13.83 16) \lvec(13.86 15) \lvec(13.89 14) \lvec(13.96 13) \lvec(14.08 12) 
    \lvec(14.27 11) \lvec(14.53 10) \lvec(14.87 9) \lvec(15.28 8) \lvec(15.77 7) 
    \lvec(16.33 6) \lvec(16.94 5) \lvec(17.59 4) \lvec(18.28 3) \lvec(19.00 2) 
    \lvec(19.74 1) \lvec(20.49 0) \lvec(16.49 0) \ifill f:0.5
    }
  \def\LandauPlot{%
    \begin{texdraw}
      \drawdim mm \setunitscale 3 \linewd 0.1
      \move(10 8.7) \lvec(20 8.7) \lvec(15 17.3) \lvec(10 8.7) \ifill f:0.8 
      \LandauLeadA \LandauLeadB 
      \move(0 0) \lvec(30 0) \lvec(30 26) \lvec(0 26) \lvec(0 0) 
      \move(0 8.7) \lvec(30 8.7) \move(1 9) \textref h:L v:B \htext{$\alpha_{3}=0$} 
      \move(5 0) \lvec(20 26) \move(4.5 1) \textref h:L v:B \rtext
      td:60 {$\alpha_{2}=0$} 
      \move(10 26) \lvec(25 0) \move(25.5 1) \textref h:R v:B \rtext
      td:-60 {$\alpha_{1}=0$}  
      \move(10 8.7) \fcir f:1 r:0.4 \textref h:C v:C \htext{$\otimes$} 
      \move(20 8.7) \fcir f:1 r:0.4 \textref h:C v:C \htext{$\otimes$} 
      \move(14.7 17) \lvec(15.3 17) \lvec(15.3 17.6) \lvec(14.7 17.6)
      \lvec(14.7 17) \lfill f:0 
      \move(14.7 8.4) \lvec(15.3 9) \move(14.7 9) \lvec(15.3 8.4) 
      \move(13.85 15.3) \fcir f:0 r:0.3 
      \move(16.15 15.3) \fcir f:0 r:0.3 
      \move(12.87 21.01) \lcir r:0.3 
      \move(17.13 21.01) \lcir r:0.3 
      \move(5 15) \LandauSEG \move (8 15) \LandauArrow{13 15.3} \move(8 16) \LandauArrow{16.5 20.5}
      \move(23 15) \LandauSEG \move (22 15) \LandauArrow{17 15.3} \move(22 16) \LandauArrow{13.5 20.5} 
      \move(15 4) \LandauBG \move(15 5) \LandauArrow{15 8}
      \move(2.5 4) \LandauHTG \move(4 5) \LandauArrow{9 8.1}
      \move(27.5 4) \LandauHTG \move(26 5) \LandauArrow{21 8.1}
      \move(15 22) \LandauLTG \move(15 21.2)  \LandauArrow{15 17.8} 
      \move(14.5 12) \LandauLSG
      \move(5 21) \LandauCoords
    \end{texdraw}
    }
  

  \def\DomainsSize{%
    \drawdim mm \setunitscale 0.8
    }
  \def\DomainsFrame{%
    \LandauLeadA \LandauLeadB
    \move(0 0) \lvec(30 0) \lvec(30 26) \lvec(0 26) \lvec(0 0)
    } 
  \def\DomainsAxes{%
    \move(0 8.7) \lvec(30 8.7) 
    \move(5 0) \lvec(20 26)    
    \move(10 26) \lvec(25 0)  
    }
  \def\DomainsSingularities{%
    \move(10 8.7) \fcir f:0 r:0.7 
    \move(20 8.7) \fcir f:0 r:0.7 
    \move(15 17.3) \fcir f:0 r:0.7 
    \move(15 8.7) \fcir f:0 r:0.7 
    \move(13.85 15.3) \fcir f:0 r:0.7 
    \move(16.15 15.3) \fcir f:0 r:0.7 
    \move(12.87 21.01) \lcir r:0.7 
    \move(17.13 21.01) \lcir r:0.7 
    }
  \def\DomainsLeading{
    \parbox[][25mm][c]{25mm}{%
      \centering
      \begin{texdraw}
        \DomainsSize    
        \DomainsFrame \lfill f:0.8
        \LandauLeadA \LandauLeadB      
        \DomainsAxes \DomainsSingularities
      \end{texdraw}
      }
    }
  \def\DomainsSubOne{%
    \parbox[][25mm][c]{25mm}{%
      \centering
      \begin{texdraw}
        \DomainsSize   
        \move(25 0) \lvec(10 26) \lvec(30 26) \lvec(30 0) \lvec(25 0) \ifill f:0.8
        \move(25 14) \textref h:C v:c \htext{(--)}
        \DomainsFrame \DomainsAxes \DomainsSingularities
      \end{texdraw}
      }
    }
  \def\DomainsSubTwo{%
    \parbox[][25mm][c]{25mm}{%
      \centering
      \begin{texdraw}
        \DomainsSize   
        \move(0 0) \lvec(5 0) \lvec(20 26) \lvec(0 26) \lvec(0 0) \ifill f:0.8
        \move(5 14) \textref h:C v:c \htext{(--)}
        \DomainsFrame \DomainsAxes \DomainsSingularities
      \end{texdraw}
      }
    }
  \def\DomainsSubThree{%
    \parbox[][25mm][c]{25mm}{%
      \centering
      \begin{texdraw}
        \DomainsSize   
        \move(0 0) \lvec(30 0) \lvec(30 8.7) \lvec(0 8.7) \lvec(0 0) \ifill f:0.8
        \move(15 2) \textref h:C v:c \htext{(--)}    
        \DomainsFrame \DomainsAxes \DomainsSingularities
      \end{texdraw}
      }
    }
  \def\DomainsSubOneTwo{%
    \parbox[][25mm][c]{25mm}{%
      \centering
      \begin{texdraw}
        \DomainsSize   
        \move(10 26) \lvec(15 17.3) \lvec(20 26) \lvec(10 26) \ifill f:0.8
        \DomainsFrame \DomainsAxes \DomainsSingularities
      \end{texdraw}
      }
    }
  \def\DomainsSubOneThree{%
    \parbox[][25mm][c]{25mm}{%
      \centering 
      \begin{texdraw}
        \DomainsSize   
        \move(25 0) \lvec(20 8.7) \lvec(30 8.7) \lvec(30 0) \lvec(25 0) \ifill f:0.8
        \DomainsFrame \DomainsAxes \DomainsSingularities
      \end{texdraw}
      }
    }
  \def\DomainsSubTwoThree{%
    \parbox[][25mm][c]{25mm}{%
      \centering 
      \begin{texdraw}
        \DomainsSize    
        \move(0 0) \lvec(5 0) \lvec(10 8.7) \lvec(0 8.7) \lvec(0 0) \ifill f:0.8
        \DomainsFrame \DomainsAxes \DomainsSingularities
      \end{texdraw}
      }
    }
  \def\DomainsI{%
    \parbox[][25mm][c]{25mm}{%
      \centering
      \begin{texdraw}
        \DomainsSize
        \move(10 8.7) \lvec(20 8.7) \lvec(15 17.3) \ifill f:0.8
        \DomainsFrame \DomainsAxes \DomainsSingularities
      \end{texdraw}
      }
    }
  \def\DomainsR{%
    \parbox[][25mm][c]{25mm}{%
      \centering 
      \begin{texdraw}
        \DomainsSize   
        \move(25 0) \lvec(20 8.7) \lvec(30 8.7) \lvec(30 0) \lvec(25 0) \ifill f:0.8
        \move(0 0) \lvec(5 0) \lvec(10 8.7) \lvec(0 8.7) \lvec(0 0) \ifill f:0.8
        \DomainsFrame \DomainsAxes \DomainsSingularities
      \end{texdraw}
      }
    }
  \def\DomainsIR{%
    \parbox[][25mm][c]{25mm}{%
      \centering 
      \begin{texdraw}
        \DomainsSize
        \move(10 8.7) \lvec(20 8.7) \lvec(15 17.3) \ifill f:0.8
        \move(25 0) \lvec(20 8.7) \lvec(30 8.7) \lvec(30 0) \lvec(25 0) \ifill f:0.8
        \move(0 0) \lvec(5 0) \lvec(10 8.7) \lvec(0 8.7) \lvec(0 0) \ifill f:0.8
        \move(3 4) \textref h:C v:C \htext{(--)}
        \move(27 4) \textref h:C v:C \htext{(--)}
        \DomainsFrame \DomainsAxes \DomainsSingularities
      \end{texdraw}
      }
    }

  \def\CutGraph{%
    \parbox[][5.5cm][c]{10cm}{%
      \begin{center}
        \begin{texdraw}
          \drawdim mm \setunitscale 1.5 
          \move(0 0) \arrowheadtype t:F \arrowheadsize l:2.5 w:1.33
          \linewd 0.1 \lpatt(1.2 1) \lcir r:10 
          \lpatt() \linewd 0.15 \larc r:10 sd:67.5 ed:112.5 \larc r:10 sd:157.5 ed:202.5
          \larc r:10 sd:247.5 ed:292.5 \larc r:10 sd:337.5 ed:382.5 
          \move(9.24 3.83) \fcir f:0 r:0.5 \lvec(13.86 5.74) 
          \move(3.83 9.24) \fcir f:0 r:0.5 \lvec(5.74 13.86)
          \move(-3.83 9.24) \fcir f:0 r:0.5 \lvec(-5.75 13.86)
          \move(-9.24 3.83) \fcir f:0 r:0.5 \lvec(-13.86 5.75)
          \move(-9.24 -3.83) \fcir f:0 r:0.5 \lvec(-13.86 -5.75)
          \move(-3.83 -9.24) \fcir f:0 r:0.5 \lvec(-5.74 -13.86)
          \move(3.83 -9.24) \fcir f:0 r:0.5  \lvec(5.74 -13.86)
          \move(9.24 -3.83) \fcir f:0 r:0.5  \lvec(13.86 -5.74)
          \move(0 10) \fcir f:1 r:0.6 \move(0 -10) \fcir f:1 r:0.6 
          \move(2 10.5) \textref h:L v:B \htext{$q_{i}$} 
          \move(2 9.2) \textref h:L v:T \htext{$i$} 
          \move(2 -10.7) \textref h:L v:T \htext{$q_{j}$} 
          \move(2 -9.2) \textref h:L v:B \htext{$j$} 
          \move(0 0) \linewd 0.1 \larc r:5 sd:-45 ed:224 
          \move(3.66 -3.41) \arrowheadtype t:F \avec(3.54 -3.54) 
          \linewd 0.15 \move(0 5) \fcir f:1 r:0.6 
          \move(2 0) \textref h:L v:B \htext{$k$} 
          \setgray 0.5 \linewd 0.3 \move(0 17) \lvec(0 -17) \linewd 0.15 \setgray 0 
          \move(2 9.8) \arrowheadtype t:F \avec(3 9.8) 
          \move(1 -10) \arrowheadtype t:F \avec(0.5 -10) 
          \move(-10 15) \textref h:R v:B \htext{$G_{1}^{[ij]}$}  
          \move(10 15) \textref h:L v:B \htext{$G_{2}^{[ij]}$} 
          \move(-15 0) \textref h:R v:C 
          \htext{$\bigl(\sum\limits_{v}(\pm)P_{v}\bigr)_{G_{1}^{[ij]}}\left\{\rule{0pt}{2.5cm}\right.$} 
          \move(15 0) \textref h:L v:C 
          \htext{$\left.\rule{0pt}{2.5cm}\right\}\bigl(\sum\limits_{v'}(\pm)P_{v'}\bigr)_{G_{2}^{[ij]}}$} 
        \end{texdraw}
      \end{center}
      }
    }  

 \def\TransitivGraph{%
    \parbox[][5.5cm][c]{10cm}{%
      \begin{center}
        \begin{texdraw}
          \drawdim mm \setunitscale 1.5 
          \move(0 0) \arrowheadtype t:F \arrowheadsize l:2.5 w:1.33
          \linewd 0.1 \lpatt(1.2 1) \lcir r:10 
          \lpatt() \linewd 0.15 \larc r:10 sd:67.5 ed:112.5 \larc r:10 sd:157.5 ed:202.5
          \larc r:10 sd:247.5 ed:292.5 \larc r:10 sd:337.5 ed:382.5 
          \move(9.24 3.83) \fcir f:0 r:0.5 \lvec(13.86 5.74) 
          \move(3.83 9.24) \fcir f:0 r:0.5 \lvec(5.74 13.86)
          \move(-3.83 9.24) \fcir f:0 r:0.5 \lvec(-5.75 13.86)
          \move(-9.24 3.83) \fcir f:0 r:0.5 \lvec(-13.86 5.75)
          \move(-9.24 -3.83) \fcir f:0 r:0.5 \lvec(-13.86 -5.75)
          \move(-3.83 -9.24) \fcir f:0 r:0.5 \lvec(-5.74 -13.86)
          \move(3.83 -9.24) \fcir f:0 r:0.5  \lvec(5.74 -13.86)
          \move(9.24 -3.83) \fcir f:0 r:0.5  \lvec(13.86 -5.74)
          \move(0 10) \fcir f:1 r:0.6 \move(0 -10) \fcir f:1 r:0.6
          \move(-10 0) \fcir f:1 r:0.6 
          \setgray 0.5
          \linewd 0.3 \move(0 17) \lvec(0 -17) 
          \move(-17 7) \lvec(7 -17)
          \move(-17 -7) \lvec(7 17)
          \setgray 0
          \linewd 0.15
          \move(2 8) \textref h:L v:T \htext{$i$} 
          \move(2 -8) \textref h:L v:B \htext{$j$} 
          \move(-8 0)  \textref h:L v:C \htext{$k$} 
          \move(2 9.8) \arrowheadtype t:F \avec(3 9.8) 
          \move(1 -10) \arrowheadtype t:F \avec(0.5 -10) 
          \move(-10 2) \arrowheadtype t:F \avec(-9.8 3) 
          \textref h:C v:C \rtext td:45 (-15 -11) {$\left\{\rule{0pt}{1.8cm}\right.$} 
          \textref h:C v:C \rtext td:-45 (-15 11) {$\left\{\rule{0pt}{1.8cm}\right.$} 
          \textref h:R v:C \rtext td:0 (-17 -12){$\bigl(\sum\limits_{v'}(\pm)P_{v'}\bigr)_{G_{2}^{[jk]}}$}
          \textref h:R v:C \rtext td:0 (-17 12){$\bigl(\sum\limits_{v'}(\pm)P_{v'}\bigr)_{G_{2}^{[ik]}}$}
          \move(15 0) \textref h:L v:C 
          \htext{$\left.\rule{0pt}{2.5cm}\right\}\bigl(\sum\limits_{v'}(\pm)P_{v'}\bigr)_{G_{2}^{[ij]}}$} 
        \end{texdraw}
      \end{center}
      }
    }

  \def\RotLabel#1#2{%
    \parbox[b][#2][b]{2ex}{%
      \begin{texdraw}
        \drawdim mm \setunitscale 1
        \move(0 0) \lvec(1 0) \lvec(0 1) \ifill f:1 
        \move(0 0) \textref h:L v:T \vtext{#1}
      \end{texdraw}
      }
    }
 

\maketitle

\begin{abstract}
  A Lorentz-covariant regularization scheme for effective field theories
  with an arbitrary number of propagating heavy and light particles is
  given.  This regularization scheme leaves the low-energy
  analytic structure of Greens functions intact and preserves all the 
  symmetries of the underlying Lagrangian.  The power
  divergences of regularized loop integrals are controlled by the
  low-energy kinematic variables.   Simple diagrammatic rules are
  derived for the regularization of arbitrary 
  one-loop graphs and the generalization to higher loops is discussed.
\end{abstract}

\textbf{PACS:}{11.10.-z, 11.15.Bt, 12.39.Hg, 12.39.Fe}

\section{Introduction}

Effective Field Theory (EFT) is a consistent framework for
calculating Greens functions below a certain scale
$\Lambda_{\text{EFT}}$ when the fundamental theory is 
unsolvable.  It is based on the principle that a physical 
amplitude can be calculated from the most general Lagrangian
formulated with the relevant degrees of freedom and consistent 
with the symmetries of the fundamental theory.  EFT then provides a
systematic method of organizing an expansion in
$p/\Lambda_{\text{EFT}}$ where $p$ is a generic  kinematic variable 
much smaller than $\Lambda_{\text{EFT}}$.  This expansion requires
a regularization scheme which preserves the power counting in the
small kinematic variables.

EFT has been applied successfully to
the calculation of Greens functions that involve strongly interacting
particles at low-energies \cite{w96}.  One example is chiral
perturbation theory ($\chi$PT) \cite{w79} where only 
quasi-Goldstone bosons of generic mass $m$ are present and
\emph{all} the kinematic 
variables are small relative to $\Lambda_\chi\approx 1$GeV.  
Dimensional regularization can be thought of as a {\it natural}
regularization scheme for $\chi$PT, because dimensionally regularized
loop integrals can
only yield terms that involve powers of the small kinematic
variables such as $m/\Lambda_\chi$.\footnote{In contrast to cutoff regularization  which
  introduces powers of the cutoff.}  Hence,  
dimensional regularization is a preferred 
regularization scheme because it allows the assignment of a chiral
power to any Feynman graph through a na\"{\i}ve dimensional
analysis while preserving the symmetries of the problem.

In the presence of heavy particles with generic masses denoted by
$M\gtrsim\Lambda_{\text{EFT}}$, 
na\"{\i}ve dimensional analysis is no longer valid because the
loop integrals also yield terms of the form $M/\Lambda_{\text{EFT}}$ which spoil
power counting \cite{gas88}.

For a single heavy particle at low energies, the dependence on $M$ can
be removed from the propagators by expanding the EFT Lagrangian in
inverse powers of 
the heavy mass as is done in 
heavy quark effective field theory and heavy baryon chiral perturbation
theory \cite{hbeft}.  The $1/M$ expansion yields loop integrals
  that satisfy low-energy power counting.  However, the $1/M$
expansion does not always reproduce 
the correct low-energy analytic structure.  An
example of this is the scalar form-factor of the nucleon 
where the $1/M$ expansion fails to reproduce the anomalous
threshold \cite{meissner}.  Therefore, the $1/M$ expansion near the
  anomalous threshold will fail to converge and an infinite number of
  terms must be re-summed.  This re-summation restores manifest
  Lorentz-invariance.  The retention of manifest Lorentz-invariance
  from the beginning avoids these problems.

\begin{figure}[tbp]
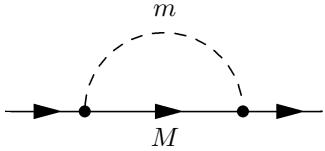

  \begin{center}
    \SelfenergyIntro
    \caption{Contribution to the nucleon self energy.}
    \label{fig:selfenergyintro}
  \end{center}
\end{figure}

A manifestly Lorentz-invariant regularization scheme which preserves the
low-energy analytic structure without spoiling power counting is
desirable.  In the presence of a single heavy particle, such a
regularization scheme called infrared regularization was devised for
one-loop Feynman graphs by 
Becher and Leutwyler~\cite{bl} building on work by Ellis and Tang~\cite{et}.
The underlying idea of infrared regularization is the separation of
power-counting violating terms from the
dimensionally regularized loop integrals, and their absorption into the
low-energy constants (LECs) of the EFT.  The power-counting violating
terms are proportional to fractional powers of $M$ which involve
non-integer values of the dimension $d$.
For example, for the nucleon self-energy graph of Fig.~\ref{fig:selfenergyintro}, the
infrared regularization separates out terms of the form $M^{d-4}/(d-4)$.
This separation is achieved by extending
the integration domain of the Feynman parameters appearing in Greens
functions.  Inspired by this work, a 
regularization of one-loop graphs with two propagating
heavy particles was recently developed in Ref.~\cite{glps} and
referred to there as EFT dimensional regularization (EFTDR).

In this paper, a consistent, Lorentz-invariant, natural regularization
scheme is presented which generalizes EFTDR to an arbitrary number of heavy
and light particles and to higher loops.  This regularization
systematically separates out all terms involving fractional powers of
$M$ implementing a low-energy power counting scheme.  To one-loop order, it
is proven that the low-energy analytic structure of the Greens
functions is left \emph{intact} by this procedure such that the
regularization is valid throughout the low-energy region.  It is demonstrated
in the appendix that this general 
formalism reproduces the one and two heavy particle sectors as first
derived in Refs.~\cite{bl,glps}. 

In Sect.~\ref{analytic}, the analytic
properties of Feynman graphs are reviewed employing the Landau
equations.  It is shown that the singularities of Feynman integrals 
can be classified according to subgraphs of the original graph.  In
Sect.~\ref{eftdr}, the regularization scheme is introduced.  It 
is shown that the fractional powers of $M$ can be associated with
subgraphs of the original graph and can be systematically
separated out.  To one-loop, simple diagrammatic rules are given to
achieve this separation.  The application of
the formalism to multi-loop graphs is discussed.  The results are
summarized in Sect.~\ref{summary}.

\section{Analytic structure of Feynman graphs}\label{analytic}

Consider a Feynman graph $G$
with $I$ scalar internal lines and $L$ loops in $d$ space-time dimensions,
\begin{equation}\label{feynman}
  I_G= \mu^{L(4-d)} \int\prod\limits^{L}_{\ell=1}\frac{\D^d k_\ell}{(2\pi)^d}
  \prod\limits^{I}_{i=1}\frac{\I}{q_i^2-m_i^2+\I \epsilon} \:,
\end{equation}
where $\mu$ is the renormalization scale of dimensional regularization.
The momentum $q_{i}^{\mu}$ of each internal line $i$ is a linear combination
of the loop momenta $k_\ell^{\mu}$ and the external momenta $P_v^{\mu}$.
The Feynman graph thus depends on the particle masses $m_i$ and a minimal
set of scalar combinations $P_v \cdot P_{v'}$ of the external momenta.
In a general EFT, fermion propagators and derivative couplings 
give rise to a momentum-dependent numerator in Eq.~\eqref{feynman}.   
Since the analytic properties of a Feynman graph only depend on 
the structure of the denominator as will be shown below, momentum
dependent numerators can be neglected for the purpose of this paper.
This simplification is not a limitation of the formalism as Feynman graphs 
involving momentum dependent numerators can always be decomposed into 
a combination of integrals of the form given by Eq.~\eqref{feynman}. 

The propagators of Eq.~\eqref{feynman} can be combined with Feynman
integrals to yield
\begin{equation}\label{feynmanrep}
  I_G=
  \I^{I} \Gamma(I) \mu^{L(4-d)}
  \int_{-\infty}^{\infty} \prod\limits_{k=1}^{I} \Bigl[\D\alpha_k\:\theta(\alpha_k)\Bigr]
  \int\prod\limits_{\ell=1}^{L}\frac{\D^d k_\ell}{(2\pi)^d}
  \frac{\delta(1-\sum\limits_{j=1}^{I} \alpha_j)}
  {\left[\sum\limits_{i=1}^{I}\alpha_{i}^{}(q_{i}^{2}-m_{i}^{2}) + \I \epsilon\right]^I}\:,
\end{equation}
where the $\theta$-functions have been introduced for later convenience.
In general, singularities of $I_G$ will arise when the contour of
integration gets pinched between two or more poles of the integrand
(pinch singularities) or between poles of the integrand and endpoints
of the integration contour (endpoint singularities).  In either case, 
the integration contour cannot be distorted away from the pole of the
integrand resulting in a singularity of the integral itself.

A useful tool for investigating analytic properties of Feynman graphs 
is given by the set of Landau equations.\footnote{See
  Refs.~\cite{polk,Itzykson} for a review.}
They constitute a necessary but not sufficient condition for the
occurrence of singularities in $I_G$ and their solution determines
the locations of the singularities in the space of masses and
external momenta as well as their locations in the space of 
integration variables $\alpha_{i}$.
In the representation \eqref{feynmanrep} of the Feynman graph,
the Landau equations for $I_G$ are given by
\begin{alignat}{2}
  &\text{for each $i$:} &\quad
  &\text{either}\quad \alpha_{i} = 0 
  \quad\text{or}\quad q_{i}^2=m_{i}^{2}\:,\label{landau2}\\
  &\text{for each $\ell$, $\mu$:} &\quad
  &\sum\limits_{i=1}^{I}\alpha_{i}^{}
  \frac{\partial q_{i}^{2}}{\partial k_{\ell}^{\mu}}=0\:,\label{landau1}
\end{alignat}
  where the $\alpha_{i}$'s must be normalized to satisfy the
$\delta$-function constraint,
\begin{equation} \label{normalization}
  \sum_{i=1}^{I} \alpha_{i} =1\:.
\end{equation}
An immediate consequence of Eq.~\eqref{landau2} is the vanishing of
the denominator in Eq.~\eqref{feynmanrep},
$  \sum_{i=1}^{I}\alpha_{i}^{}(q_i^2-m_i^2)=0\:.$
Eq.~\eqref{landau2} means that an internal
particle $i$ of the Feynman graph $G$ is either 
on-shell or the Feynman parameter $\alpha_i$ of line $i$ is zero.
If $\alpha_i=0$, line $i$ does not contribute to the remaining 
Landau equations \eqref{landau1}. In this case, the set of Landau equations
for $G$ is identical to the Landau equations for the subgraph $g$ of $G$
that is obtained by contracting line $i$ to a point.\footnote{It is
  important to emphasize that the subgraphs thus obtained are not
  Feynman graphs and do not represent physical amplitudes.
  Henceforth, these subgraphs will be represented by lower-case letters.}

A solution to the Landau equations where all the $\alpha_j$'s are non-zero,
\textit{i.e.}, all internal lines are on-shell, is called the 
\emph{leading singularity} of $G$;
singularities with $n$ internal lines off-shell are
called \emph{(sub)$^{n}$-leading singularities}.
It follows that every (sub)$^{n}$-leading singularity of a Feynman graph $G$ 
can be represented by the leading singularity of a subgraph $g$ where
the corresponding $n$ internal lines 
of the original graph have been contracted.
The complete set of singularities of a Feynman graph $G$ is then given by
the leading singularities of each of its subgraphs.

For one-loop graphs, a more practical formulation of the Landau
equations can be derived. Routing all internal momenta $q_{i}^{\mu}$
to flow around the loop, Eq.~\eqref{landau1} takes on the form
\begin{equation}
  \label{eq:landau3}
  \text{for each $i$ with $\alpha_{i}\neq 0$:} \qquad
  \sum_{j=1}^{I} \alpha_{j}^{} \: q_{j}^{\mu} = 0\:,
\end{equation}
where only uncontracted lines $j$ contribute. Multiplying this
equation by $q_{i \mu}$ for each uncontracted line $i$ yields a 
set of linear equations
\begin{equation}
  \label{eq:landau4}
  \sum_{j=1}^{I} (q_{i}\cdot q_{j}) \: \alpha_{j} = 0\:.
\end{equation}
Using the on-shell condition \eqref{landau2}, the matrix $(q_{i}\cdot
q_{j})$ can be re-written as 
\begin{equation}
  \label{eq:landau5}
  \Omega_{ij} = (q_{i}\cdot q_{j}) = \frac{m_{i}^{2}+m_{j}^{2}-(q_{i}-q_{j})^{2}}{2} \:,
\end{equation}
which is now independent of the loop momentum $k^{\mu}$.
The location of the singularity in the space of masses and external
momenta is determined from the condition 
\begin{equation}
  \label{eq:landau6}
  \det \Omega_{ij} = 0\:,
\end{equation}
while the corresponding location in Feynman parameter space is obtained by
solving the zero-eigenvalue problem \eqref{eq:landau4} and normalizing 
the $\alpha_{i}$'s according to Eq.~\eqref{normalization}. A singularity
does not arise on the physical sheet when the solution in parameter
space lies outside the domain of integration of the Feynman graph.
\begin{figure}[htb]
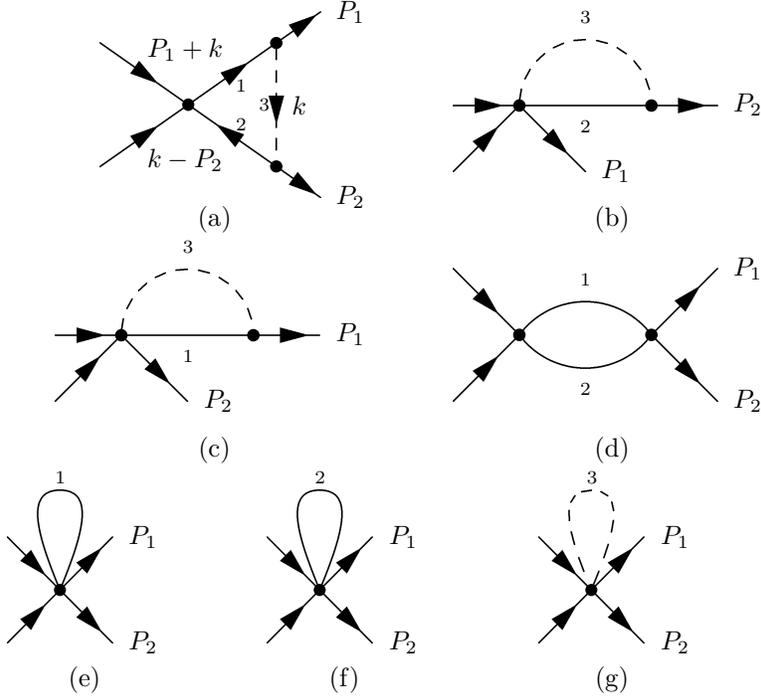

  \begin{center}
    \begin{tabular}[c]{c@{\hspace{40pt}}cc}
      \multicolumn{2}{c}{\TriangleGraph}    & \SelfenergyGraphA \\
      \multicolumn{2}{c}{(a)}               & (b) \\
      \multicolumn{2}{c}{\SelfenergyGraphB} & \BubbleGraph \\
      \multicolumn{2}{c}{(c)}                & (d) \\
      \HeavyTadpoleGraph{1} & \HeavyTadpoleGraph{2} &
      \LightTadpoleGraph\\
      (e) & (f) & (g)
    \end{tabular}
    \caption{Triangle graph (a) and all its subgraphs (b)--(g).}
    \label{fig:triangle}
  \end{center}
\end{figure}

\subsection{An example: the triangle graph}
As an illustration consider the triangle graph shown in Fig.~\ref{fig:triangle}(a)
where the solid lines represent heavy particles of mass $M$ and the
dashed line represents a light particle of mass $m$.
The corresponding subgraphs are shown in
Figs.~\ref{fig:triangle}(b)-(g).

\begin{figure}[htbp]
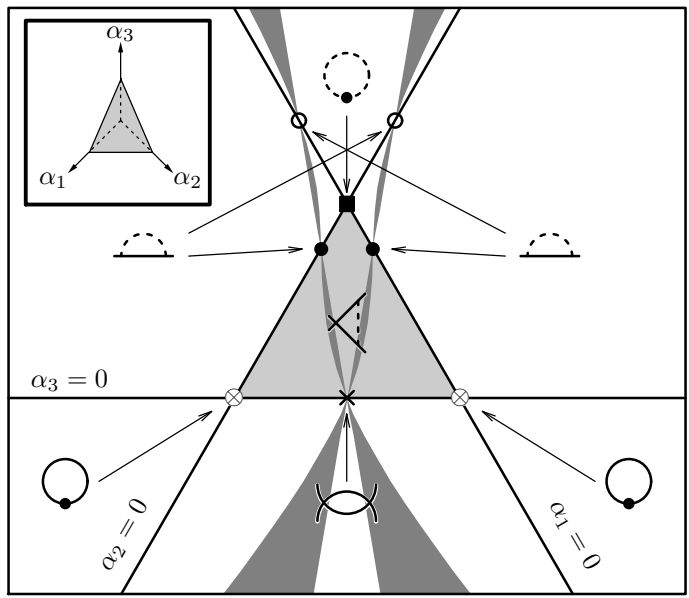

  \begin{center}
    \LandauPlot
    \caption{Integration domain and location of singularities of the triangle graph.}
    \label{fig:Landau}
  \end{center}
\end{figure}

For the triangle graph, the $\delta$-function appearing in Eq.~\eqref{feynmanrep} constrains 
the domain of integration to a plane in 
the $(\alpha_1,\alpha_2,\alpha_3)$-space. Fig.~\ref{fig:Landau} shows the
integration domain and the locations of the singularities in this
plane. The integration domain (light shaded area) 
is bounded by the lines $\alpha_{i}=0$.
A (sub)$^{1}$-leading singularity corresponding to
the subgraph where 
line $i$ is contracted lies on the line
$\alpha_{i}=0$. A (sub)$^{2}$-leading singularity represented by 
the subgraphs where lines $i$ and $j$ are contracted lies
at the intersection of the lines $\alpha_{i}=0$ and $\alpha_{j}=0$
which is a corner of the integration domain. Finally, the Feynman
parameters that correspond to the leading singularity form a surface
given by the dark shaded area in Fig.~\ref{fig:Landau} which penetrates
the integration boundaries at the location of the (sub)$^{1}$-leading 
singularities.  In Feynman parameter space, the leading
singularity is a pinch singularity while
all the sub-leading singularities occurring on the physical sheet
lie on the integration boundary and are therefore endpoint
singularities.

It is instructive to solve the Landau equations 
for each subgraph:
\begin{itemize}
\item \textit{ $(\text{Sub})^{2}$-leading singularities.}
  For the heavy-particle tadpole subgraph Fig.~\ref{fig:triangle}(e)
  with lines 2 and 3 contracted,
  Eqs.~\eqref{eq:landau4} and \eqref{eq:landau5} yield immediately
  \begin{equation} \label{triangle_tadpole}
    \begin{split}
      &\alpha_{1}\neq 0\:, \quad \alpha_{2}=\alpha_{3}=0\:,\\
      &\alpha_{1} M^{2} = 0 \quad \Rightarrow M^2 = 0\:.
    \end{split}
  \end{equation}
  Similarly, the light-particle tadpole Fig.~\ref{fig:triangle}(g)
  yields a singularity at $m^2=0$. In Fig.~\ref{fig:Landau}, the
  heavy-particle tadpole singularities are denoted by $\otimes$,
  the light-particle tadpole by~$\blacksquare$.

\item \textit{$(\text{Sub})^{1}$-leading singularities}. For the 
  self-energy subgraph in Fig.~\ref{fig:triangle}(b)  condition
  \eqref{eq:landau6} reads
  \begin{equation} \label{triangle_selfenergy}
    \begin{split}
      &\alpha_{1}=0\:, \quad\alpha_{2}\neq 0\:,\quad\alpha_{3}\neq 0\:,\\
      &\begin{vmatrix}
        M^2 & \frac{M^2+m^2-P_{2}^{2}}{2} \\
        \frac{M^2+m^2-P_{2}^{2}}{2} & m^2
      \end{vmatrix}
      = \left[ P_{2}^{2}-(M+m)^2\right] \left[P_{2}^{2}-(M-m)^2\right]=0\:.
    \end{split}
  \end{equation}
  The corresponding zero-eigenvectors are easily
  determined. The singularities are located at
  \begin{alignat}{2}
    &P_{2}^{2} = (M+m)^2  &\quad  &\text{for}\quad 
    \alpha_{2}=\frac{m}{M+m}\:,\: \alpha_{3}=\frac{M}{M+m}\:,
    \label{HL_threshold}\\
    &P_{2}^{2} = (M-m)^2  &\quad  &\text{for}\quad 
    \alpha_{2}=-\frac{m}{M-m}\:,\: \alpha_{3}=\frac{M}{M-m}\:.
    \label{HL_pseudothreshold}
  \end{alignat}
  Solution \eqref{HL_threshold} (denoted by $\bullet$ in Fig.~\ref{fig:Landau}) 
  is recognized as the physical heavy-light-threshold, while the pseudo-threshold 
  \eqref{HL_pseudothreshold} (denoted by $\circ$) does not appear as a singularity 
  on the physical sheet since it lies outside ($\alpha_{2}<0$) 
  the domain of integration 
  of the Feynman graph \eqref{feynmanrep}.
  Similarly, the thresholds for the  bubble subgraph in
  Fig.~\ref{fig:triangle}(d) are
  \begin{alignat}{2}
    &P^{2} = 4M^{2}\:  &\quad  &\text{for}\quad
    \alpha_{1}=\alpha_{2}=\frac{1}{2}\:,\quad \alpha_{3}=0\:, \qquad 
    (\mathbf{\times}\:\text{in Fig.~\ref{fig:Landau}}); 
    \label{HH_threshold}\\
    &P^{2} = 0\: &\quad &\text{not on the physical sheet\footnotemark}.
  \end{alignat}
  \footnotetext{The normalization condition \eqref{normalization} cannot 
    be fulfilled.}  
\item \textit{Leading Singularity.}
  The leading singularity of the triangle graph, corresponding to 
  $\alpha_{i}\neq 0$ $(i=1,2,3)$, is a manifold rather than a point 
  in the space of independent external momenta $P_{1}^{2}$,
  $P_{2}^{2}$, and $P^{2}=(P_{1}+P_{2})^2$ and satisfies
  \begin{equation}
    \label{eq:triangle_leading}
    \begin{vmatrix}
      M^2 & \frac{2 M^{2} -P^{2}}{2} &
      \frac{M^{2}+m^{2}-P_{1}^{2}}{2} \\
      \frac{2M^{2}-P^{2}}{2} & M^2 &
      \frac{M^{2}+m^{2}-P_{2}^{2}}{2} \\
      \frac{M^{2}+m^{2}-P_{1}^{2}}{2} &
      \frac{M^{2}+m^{2}-P_{2}^{2}}{2} & m^2
    \end{vmatrix}
    = 0\:.
  \end{equation}
  To this manifold in momentum space corresponds a manifold in Feynman 
  parameter space given by the dark shaded area in Fig.~\ref{fig:Landau}.
\end{itemize}

\section{EFT Dimensional Regularization}\label{eftdr}

In this section, a natural regularization scheme (EFTDR) for an EFT applicable 
below a scale $\Lambda_{\text{EFT}}$ is established.  It is the
low-energy kinematic variables that control the divergences of the loop
integrals regularized using EFTDR.  
The low-energy domain is defined by the relations
\begin{equation}
  \begin{split}
    |P_v\cdot P_{v'} - M^2| &\ll \Lambda_{\text{EFT}}^2\:,\\
    P_v\cdot p_{v'},\: p_v\cdot p_{v'},\: m^2 &\ll \Lambda_{\text{EFT}}^2\:,\\
    M &\gtrsim \Lambda_{\text{EFT}}\:,
  \end{split}
\end{equation}
where $P_{v}^{\mu}$ and $p_{v}^{\mu}$ are 
the heavy and light particle external momenta respectively, while $M$ and $m$ are
generic heavy and light particle masses respectively.

EFTDR is implemented by consistently separating a loop integral
$I_{G}$ corresponding to a graph $G$ into two parts,
\begin{align}
  \label{eq:Decomposition}
  I_{G} = \bar{I}_{G} + R_{G}\:,
\end{align}
where the entire low-energy analytic structure of $I_G$ is collected
into $\bar{I}_{G}$ while all the terms multiplied by factors of the form
$M^{\ell d-n}$ (referred to as fractional powers of
$M$ for fractional space-time dimension $d$) where $\ell$ and $n$ are
integers, are collected into $R_G$, 
the \emph{regular part}.\footnote{The 
  terminology is taken from Ref.~\cite{bl} and refers to the fact that
  \emph{to one-loop}, $R_{G}$ is analytic in the low-energy region.}  The
analytic structure of $\bar{I}_{G}$ and $I_G$ are {\it identical}
in the low-energy domain such that $R_G$ can be accounted
for through a renormalization of the low-energy constants of the EFT.
The separation \eqref{eq:Decomposition} will be achieved by
systematically separating out the subgraphs representing
singularities that lie outside of the low-energy region; these
subgraphs will be referred to as \emph{regular subgraphs}.

In the example of the triangle graph, the only singularities that lie
outside of the low-energy region are the $M=0$ singularities
represented by Figs.~\ref{fig:triangle}(e)-(f).  Since a factor $M^{\ell d-n}$
will be singular at $M=0$ for appropriate $d$, separating out the $M=0$
singularity will ensure that the terms multiplied
by fractional powers of $M$ are separated out.

\begin{figure}[tbp]
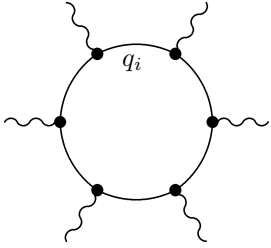

  \begin{center}
    \SoftHeavyRingGraph{6}
    \caption{A regular subgraph composed of heavy particle propagators 
      and low-energy momentum insertions represented by wiggly lines.}
    \label{fig:SoftHeavyRing}
  \end{center}
\end{figure}

Generally, subgraphs containing only heavy
particle propagators and low-energy momentum insertions are regular subgraphs that
give rise to fractional powers of $M$.  Indeed,
the Landau equations \eqref{eq:landau4} for the regular
subgraph shown in Fig.~\ref{fig:SoftHeavyRing} are 
\begin{eqnarray}\label{ringsln}
M^2\left[1-\sum\limits_{j=1}^I\frac{1}{2M^2}(q_i-q_j)^2\alpha_j\right]
\approx M^2= 0\:.
\end{eqnarray}
Therefore, these subgraphs cannot have singularities in the low-energy
region and belong in $R_G$.

\begin{figure}[htbp]
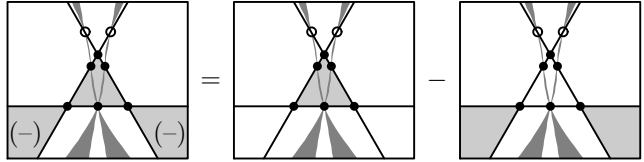

  \begin{center}
    \DomainsIR = \DomainsI $-$ \DomainsR
    \caption{Decomposition of the integration contour for the triangle graph}
    \label{fig:TriangleDecomposition}
  \end{center}
\end{figure}

The subgraphs that lead to fractional powers of $M$ can be isolated by 
extending the Feynman parameter domain of integration in
Eq.~(\ref{feynmanrep}) 
where the $\theta(\alpha_k)$ define the integral
boundaries.  The product of $\theta$'s can be rewritten:
\begin{multline}\label{thetaprod2}
  \delta(1-\sum\limits_{k=1}^I \alpha_k)
  \prod\limits^I_{i=1}[1-\theta(-\alpha_i)]\\
  = \delta(1-\sum\limits_{k=1}^I \alpha_k) 
  \biggl\{ 1 + \sum\limits^I_{i=1}[-\theta(-\alpha_i)] \\
  + \sum\limits^I_{\substack{i,j=1\\i<j}} 
  [-\theta(-\alpha_i)] [-\theta(-\alpha_j)] 
  + \cdots\biggr \}\:.
\end{multline}
It is noted that each $\theta(-\alpha_i)$ contains the point
$\alpha_i=0$ which, as discussed in the previous section, represents a
contracted line in $G$ that leads to a subgraph of $G$.  In light of
this, each term in
Eq.~(\ref{thetaprod2}) defines a domain of integration $\Delta_{g}$,
\begin{align}\label{domains}
  \Delta_{g}= 
   \delta(1-\sum\limits_{k=1}^I \alpha_k)\prod\limits_{i\notin g}
   \left[-\theta(-\alpha_{i})\right]\:, 
\end{align}
which can be associated with a particular subgraph $g$ of
$G$.\footnote{The subscript $g$ represents a set of indices labeling
  the internal lines of the subgraph $g$}
$R_{G}$ is then defined as
\begin{eqnarray}\label{regular}
  R_{G}=\I^I\Gamma(I)\mu^{L(4-d)}\int_{-\infty}^\infty \prod\limits^I_{k=1} \D \alpha_k\: 
  {\mathcal{D}}_{R_{G}} \int\prod\limits^{L}_{\ell=1}\frac{\D^d k_\ell}{(2\pi)^d}
  \biggl[\sum\limits^I_{i=1} \alpha_i^{}(q^2_i-m^2_i)+\I\epsilon\biggr]^{-I},
\end{eqnarray}
where
\begin{eqnarray} \label{domain_function}
  {\cal{D}}_{R_{G}} = \!\!\! \sum_{{\text{regular $g$'s}}} \!\!\! \Delta_g.
\end{eqnarray}
It is instructive to discuss the triangle graph $I_{3}$ as an
illustration of the 
method.  Tab.~\ref{tab:domains} shows the association of each term in
the $\theta$-expansion with the corresponding subgraph and domain of
integration.  As mentioned above, the only regular subgraphs are the heavy
particle tadpoles located at
$(\alpha_{1},\alpha_{2},\alpha_{3})=(1,0,0)$ and $(0,1,0)$ in
Fig~\ref{fig:Landau}.  The  
integration domains in Eq.~\eqref{thetaprod2} that contain these
points as endpoints are: 
$\theta(-\alpha_{2})\theta(-\alpha_{3})$ and
$\theta(-\alpha_{1})\theta(-\alpha_{3})$.  These integration
domains do not include any of 
the low-energy singularities; this implies that $R_{3}$ is analytic in the
low-energy domain.  $\bar{I}_{3}$ is then given by
\begin{equation}\label{ibartriangle}
  \bar{I}_{3}=I_{3}-R_{3}\:,
\end{equation}
and possesses the same low-energy analytic structure in the small
kinematical variables as $I_{3}$ but does not contain any fractional
powers of $M$.  This is evident by looking at the integration domains
corresponding to Eq.~\eqref{ibartriangle} and
drawn in Fig.~\ref{fig:TriangleDecomposition}.

\begin{table}[htbp]
  \begin{center}
    \begin{tabular}[c]{||c|c|l|c|c||}
      \hline\hline
      & subgraph $g$ & domain function $\Delta_{g}$ & integration domain &
      \\ \hline\hline
      \raisebox{-0.5cm}{\RotLabel{leading}{1em}}
      &\TriangleGraph\rule[-39pt]{0pt}{82pt} & $\delta(1-\sum\limits_{i=1}^{3}\alpha_{i})$  
      &\DomainsLeading 
      &
      \\ \cline{1-4}
      &\SelfenergyGraphA 
      &\parbox[]{3cm}{$-\theta(-\alpha_{1})\:$\\
        \rule{0.6cm}{0pt}$\times\delta(1-\sum\limits_{i=1}^{3}\alpha_{i})$}  
      &\DomainsSubOne  
      &      
      \\[-2pt] \cline{2-4}
      \raisebox{-1cm}{\RotLabel{(sub)$^{1}$-leading}{1em}}
      &\SelfenergyGraphB 
      &\parbox[]{3cm}{$-\theta(-\alpha_{2})\:$\\
        \rule{0.6cm}{0pt}$\times\delta(1-\sum\limits_{i=1}^{3}\alpha_{i})$}  
      &\DomainsSubTwo
      &
      \\[-2pt] \cline{2-4}
      &\BubbleGraph 
      &\parbox[]{3cm}{$-\theta(-\alpha_{3})\:$\\
        \rule{0.6cm}{0pt}$\times\delta(1-\sum\limits_{i=1}^{3}\alpha_{i})$}
      &\DomainsSubThree
      &
      \\[-2pt] \cline{1-4}
      &\LightTadpoleGraph 
      &\parbox[]{3cm}{$\theta(-\alpha_{1})\:\theta(-\alpha_{2})$\\
        \rule{0.6cm}{0pt}$\times\delta(1-\sum\limits_{i=1}^{3}\alpha_{i})$} 
      &\DomainsSubOneTwo  
      &\raisebox{2cm}[0pt]{\RotLabel{subgraphs with low-energy singularities}{1em}}
      \\ \cline{2-5}
      \raisebox{-1cm}{\RotLabel{(sub)$^{2}$-leading}{1em}}
      &\HeavyTadpoleGraph{2} 
      &\parbox[]{3cm}{$\theta(-\alpha_{1})\:\theta(-\alpha_{3})\:$\\
        \rule{0.6cm}{0pt}$\times\delta(1-\sum\limits_{i=1}^{3}\alpha_{i})$}
      &\DomainsSubOneThree  
      &
      \\ \cline{2-4}
      &\HeavyTadpoleGraph{1}  
      &\parbox[]{3cm}{$\theta(-\alpha_{2})\:\theta(-\alpha_{3})\:$\\
        \rule{0.6cm}{0pt}$\times\delta(1-\sum\limits_{i=1}^{3}\alpha_{i})$} 
      &\DomainsSubTwoThree 
      &\RotLabel{regular subgraphs}{1em}
      \\\hline\hline
    \end{tabular}
    \caption{Integration domains for the triangle graph}
    \label{tab:domains}
  \end{center}
\end{table}

For a general Feynman graph $G$, a particular domain  of 
integration $\Delta_g$ associated with a subgraph $g$ will never
select a singularity associated with any subgraph of $g$.  Indeed, the 
domain of integration $\Delta_g$ will never have as an \emph{endpoint}
the location in $\alpha$-space of any singularity associated with a
subgraph of $g$.  It follows that a subgraph $\bar{g}$ that contributes to
$\bar{I}_{G}$ can never give rise to fractional powers of $M$ since they
can only arise from regular subgraphs that are subgraphs of
$\bar{g}$.  This proves that $R_G$ contains all the fractional
powers of $M$.  To one loop, it is also straightforward to show that
$R_G$ is analytic in the low-energy region.

\subsection{One-loop graphs}

For one-loop graphs, more explicit results can be
obtained.  Evidently, any subgraph of a regular subgraph, obtained by contracting
out further lines, is also regular. This gives rise to the notion of
a \emph{minimally contracted regular} (MCR) subgraph,
\textit{i.e.},  a regular  subgraph ${\mathfrak{g}}$ that is \emph{not} 
a subgraph of any other regular subgraph 
with more lines than ${\mathfrak{g}}$. In EFTs to one-loop order, there are at most 
two different MCR subgraphs of a given graph and every internal heavy-particle 
line belongs to exactly one of them (cf. App.~\ref{sec:MCRSG}).  As an 
example, the two tadpoles of Figs.~\ref{fig:triangle}(e)-(f) are the
MCR subgraphs of the triangle graph.

Since by construction different MCR subgraphs do not have common lines,
they cannot have subgraphs in common. Any regular subgraph $g$  
is thus a subgraph of exactly one MCR subgraph ${\mathfrak{g}}$. 
The domain functions \eqref{domains} of a MCR subgraph
${\mathfrak{g}}$ and all its subgraphs $g\subset{\mathfrak{g}}$ 
can be re-summed to yield
\begin{align}
  \sum\limits_{g\subseteq{\mathfrak{g}}}\Delta_{g} &= \delta(1-\sum_{k=1}^{I} \alpha_{k})\:
   \biggl\{ 1 - \sum_{i\in {\mathfrak{g}}} \theta(-\alpha_{i}) + \cdots
   \biggr\} \:
  \prod_{j\notin {\mathfrak{g}}} \bigl[-\theta(-\alpha_{j})\bigr]\\
  &=  \delta(1-\sum_{k=1}^{I} \alpha_{k})\: 
  \prod_{i\in {\mathfrak{g}}}  \theta(\alpha_{i}) \:
   \prod_{j\notin {\mathfrak{g}}}\bigl[-\theta(-\alpha_{j})\bigr] \:.
 \end{align}
The domain function ${\mathcal{D}}_{R_{G}}$ introduced in
Eq.~\eqref{domain_function} is then
\begin{equation}
  \label{eq:domain_function2}
  {\mathcal{D}}_{R_{G}} = \sum\limits_{\mathfrak{g}}
  \sum\limits_{g\subseteq{\mathfrak{g}}}\Delta_{g}\:,
\end{equation}
and the full  regular part is given as 
\begin{align}\label{regularpart}
  R_{G} &= \I^{I} \Gamma(I) \mu^{4-d} \sum_{\mathfrak{g}}
  \int_{0}^{\infty}\! \prod_{i\in {\mathfrak{g}}}\D \alpha_{i}\:
  \int_{-\infty}^{0}\! \prod_{j\notin {\mathfrak{g}}}(-\D \alpha_{j})\: 
  \delta(1-\sum_{k=1}^{I} \alpha_{k})\: \nonumber \\
  &\qquad \qquad \times \int\!\!\frac{\D^d k}{(2 \pi)^{d}}\:
  \biggl[\sum\limits_{i=1}^{I} \alpha_{i}^{} (q_{i}^{2}-m_{i}^{2})
      + \I \epsilon\biggr]^{-I}.
\end{align}
The momentum integration can be
carried out explicitly \cite{Itzykson} yielding
\begin{align}
  R_{G} = \kappa_{I}  \sum_{\mathfrak{g}}
  \int_{0}^{\infty}\!\ \prod_{i\in {\mathfrak{g}}}\D \alpha_{i}\:
  \int_{-\infty}^{0}\! \prod_{j\notin {\mathfrak{g}}}(-\D \alpha_{j})\: 
  \delta(1-\sum_{k=1}^{I} \alpha_{k})\: 
  \bigl[ C -\I \epsilon\bigr]^{d/2-I}\:, \label{thirdrep}
\end{align}
where:
\begin{equation}
  \label{eq:denominator}
  C = \sum_{i,j = 1}^{I} \alpha_{i} \Omega_{ij} \alpha_{j} \:, \qquad 
  \Omega_{ij} = \frac{m_{i}^{2}+m_{j}^{2}-(q_{i}-q_{j})^{2}}{2}.
\end{equation}
The matrix $\Omega_{ij}$ was defined in Eq.~\eqref{eq:landau5} and the
factors have been collected in 
\begin{equation}
  \label{eq:kappa}
  \kappa_{I} = (-)^{I} \frac{\I^{I+1}}{16 \pi^{2}} \frac{\Gamma(I-d/2)}{ (4
    \pi\mu^2 )^{d/2-2}} \:.
\end{equation}

A useful representation of $R_G$ with a single non-compact integral
can be derived by inserting the identity
\begin{equation}
  1 = \int_1^\infty \!\!\D \lambda \: \delta(\lambda-
  \sum\limits_{k \in {\mathfrak{g}}} \alpha_k)
\end{equation}
into Eq.~\eqref{thirdrep} 
and performing the change of variables
\begin{equation}
  \label{cov}
  \begin{alignedat}[c]{2}{2}
    \alpha_{i} &= \lambda z_{i} &\qquad i&\in  {\mathfrak{g}}\:,\\
    \alpha_{i} &= (1-\lambda) z_{i} &\qquad i&\notin {\mathfrak{g}}\:.
  \end{alignedat}
\end{equation}
This yields for $R_{G}$:
\begin{multline}
  \label{noncompactrep}
  R_{G}=-\kappa_I  \sum\limits_{\mathfrak{g}} \int_1^\infty\!\!\D \lambda \:
  \lambda^{|{\mathfrak{g}}|-1}(1- \lambda )^{I-|{\mathfrak{g}}|-1} \\ 
  \times \int_0^1 \left(\prod\limits_{i\in {\mathfrak{g}}} \D z_i\right) 
  \delta(1- \sum\limits_{k\in{\mathfrak{g}}} z_k)
  \int_0^1 \left(\prod\limits_{j\notin{\mathfrak{g}}} \D z_j\right) 
  \delta(1- \sum\limits_{k\notin{\mathfrak{g}}}z_k) \\ 
  \times\Bigg[(1-\lambda)^2 \sum\limits_{i,j\notin{\mathfrak{g}}} z_i\Omega_{ij}z_j +
  \lambda^2 \sum\limits_{i,j \in{\mathfrak{g}}}  z_i\Omega_{ij}z_j \\
  + 2\lambda(1-\lambda) \sum\limits_{\substack{i\in{\mathfrak{g}}\\j\notin{\mathfrak{g}}}} 
    z_i\Omega_{ij}z_j - \I  \epsilon \Bigg]^{d/2-I}\:,
\end{multline}
where $|{\mathfrak{g}}|$ denotes the number of internal lines
belonging to the MCR subgraph ${\mathfrak{g}}$.

In this representation, it is straightforward to prove that $R_{G}$ is 
analytic in the low-energy region.
As stated above, every internal heavy-particle line belongs to
one of the two MCR subgraphs ${\mathfrak{g}}_{1}$ or ${\mathfrak{g}}_{2}$.\footnote{%
  Here ${\mathfrak{g}}_{2}$ is understood to be empty in cases where there
  is only one MCR subgraph.}
Denoting the set of internal light-particle lines 
by $\mathfrak{l}$, the elements of the matrix appearing in 
Eq.~\eqref{eq:denominator} are 
(cf. Eq.~\eqref{eq:qiqj} in App.~\ref{sec:MCRSG})
\begin{equation}
  \Omega_{ij} =
  \begin{cases}
    {\mathcal{O}}(p^2)   &i,j\in {\mathfrak{l}} \\
    {\mathcal{O}}(p) M   &i\in{\mathfrak{l}}\:,\quad j\notin{\mathfrak{l}} \\
    M^2+{\mathcal{O}}(p^2)  &i,j\in {\mathfrak{g}}_{1} \quad \text{or} \quad i,j\in{\mathfrak{g}}_{2} \\
    -M^2+{\mathcal{O}}(p)M  &i\in{\mathfrak{g}}_{1}\:, \quad j\in{\mathfrak{g}}_{2}
  \end{cases}
  \:,
\end{equation}
yielding for the denominator of Eq.~\eqref{noncompactrep} with ${\mathfrak{g}}={\mathfrak{g}}_{1}$:
\begin{equation}   \label{mcrsg_denom}
  \begin{split}
    \sum\limits_{i,j=1}^I \alpha_i\Omega_{ij}\alpha_j &=
    (1-\lambda)^2\sum\limits_{i,j\in{\mathfrak{g}}_{2}} z_i M^{2} z_j 
    - 2\lambda (1-\lambda)\sum\limits_{\substack{i\in{\mathfrak{g}}_{1}\\j\in{\mathfrak{g}}_{2}}} z_i M^{2} z_j\\
    &\qquad \qquad + \lambda^{2} \sum\limits_{i,j\in{\mathfrak{g}}_{1}} z_i M^{2} z_j + {\mathcal{O}}(p) M\\
    &= M^2(1+\nu)^2\lambda^2\left[1-\frac{\nu}{(1+\nu)\lambda}\right]^2
    +{\mathcal{O}}(p) M ,
  \end{split}
\end{equation}
where $\nu=\sum\limits_{i\in{\mathfrak{g}}_{2}} z_i$ with $0\leq\nu\leq 1$ and
$\lambda\geq 1$.  Thus, the denominator never vanishes proving that
$R_{G}$ is always analytic for one-loop graphs with arbitrary
numbers of heavy and light particle propagators.  The integral over
$\lambda$ can be performed by expanding the square bracket in
Eq.~\eqref{mcrsg_denom}.  To leading order in this expansion, this
gives 
\begin{multline}\label{generalR}
  R_{G} \simeq
  \sum_{\mathfrak{g}}
  \frac{(-1)^{I-|{\mathfrak{g}}|-1} M^{d-2I} \kappa_I}{(d-I-1)\Gamma(|{\mathfrak{g}}|)}\\
  \times\int_0^1
  \left(\prod\limits_{i\notin{\mathfrak{g}}} \mbox{d}z_i\right) 
  \delta(1- \sum\limits_{k\notin {\mathfrak{g}}} z_k)
  (1+\nu)^{d-2I} +M{\mathcal{O}}(p)\:,
\end{multline}
where the $\lambda$-integral was performed at $d$ sufficiently small
to drop the upper boundary; $d$ can now be analytically continued to
4 space-time dimensions.  Eq.~\eqref{generalR} can be used to
calculate to leading order the regular part of $R_{3}$ keeping in mind 
that there are two MCR subgraphs for the triangle graph:
\begin{align}\label{R3}
  R_{3}=-\frac{1}{16\pi^2}\frac{\Gamma(3-d/2)}{(4\pi\mu^2)^{d/2-2}}\frac{M^{d-6}}{(d-4)(d-5)}(1+\cdots)\:.
\end{align}
Eq.~\eqref{R3} agrees with the result given in Ref.~\cite{glps}.

Eq.~\eqref{generalR} explicitly shows 
that $R_G$ is proportional to a fractional power of $M$.  This implies 
that $R_G$ and $\bar{I}_G$ transform separately under the symmetry
transformations that leave the Lagrangian invariant \cite{bl}.  This ensures that 
EFTDR preserves all the symmetries of the underlying Lagrangian.  For general
one-loop graphs, Eq.~\eqref{generalR} also shows that the
non-compact integrals appearing in $\bar{I}_G$ can be performed quite
generally.

The Taylor series expansion of $R_{G}$ in powers of $p$ is
truncated at the order of the EFT Lagrangian.  It follows that the
Taylor-polynomial of $R_{G}$ can be absorbed into the
low-energy constants of 
the EFT Lagrangian.  Once $R_{G}$ has been used to define the domain
of integration of $\bar{I}_G$, it need not be considered further.

For a general one-loop Feynman graph $G$, 
the EFT dimensional regularization scheme is summarized by 
the following simple rules:
\begin{enumerate}
  
\item Label all of the internal lines $i$ of the graph $G$ by the
  corresponding Feynman parameter $\alpha_{i}$.
  
\item Determine the minimally contracted regular subgraph(s) ${\mathfrak{g}}$ of $G$.

\item For each minimally contracted regular subgraph ${\mathfrak{g}}$, 
  integrate the Feynman parameters corresponding to internal lines of 
  ${\mathfrak{g}}$ from $0$ to $+\infty$; integrate the remaining Feynman 
  parameters from $-\infty$ to $0$ inserting a factor of -1 for every integral.
  
\item Sum these contributions for all minimally contracted subgraphs
  to obtain the regular part $R_{G}$.
  
\item The EFT dimensionally regularized Feynman integral is given as
  \begin{displaymath}
    \bar{I}_{G} = I_{G} - R_{G}\:.
  \end{displaymath}
  
\end{enumerate}

\subsection{Discussion of multi-loop graphs}\label{multiloop}

\begin{figure}[tbp]
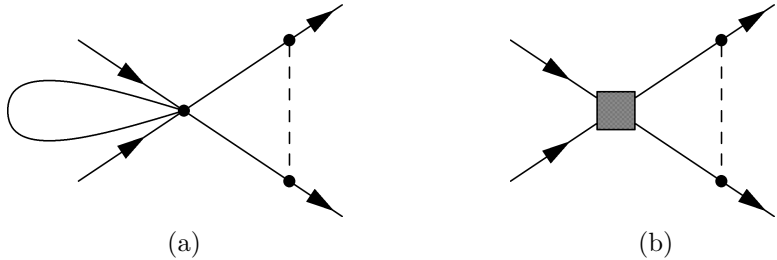

  \begin{center}
    \begin{tabular}{c@{\hspace{1cm}}c}
      \TwoLoopGraph  &\RenormTriangleGraph \\
      (a) & (b)
    \end{tabular}
    \caption{Figure~(a) shows an example of an unnested two-loop graph
      that gives rise to fractional powers of $M$ while containing
      low-energy singularities. Figure~(b) shows the triangle graph with 
      renormalized four-heavy-particle vertex.}
    \label{fig:twoloop}
  \end{center}
\end{figure}

It is emphasized that the regularization scheme developed above is not
limited to one-loop graphs. Indeed, the Landau equations
are valid to arbitrary number of  loops and allow a determination
of those subgraphs that give rise to fractional powers of $M$.
The contributions from these and only these subgraphs are collected
in $R_{G}$ and need to be separated from the dimensionally regularized
Feynman integral $I_{G}$ to obtain the EFT dimensionally regularized integral 
$\bar{I}_{G}$. The separation
\begin{equation}
  I_{G} = \bar{I}_{G}+R_{G}
\end{equation}
is again achieved by extending the domain of integration 
as explained above. 
As a novel feature at two or more loops,  subgraphs that give rise to 
fractional powers of $M$ may now contain light internal lines and
may give rise to low-energy singularities. 
An example of such a subgraph is a tadpole attached to a triangle graph 
as shown in  Fig.~\ref{fig:twoloop}(a):  the tadpole part of the graph 
gives rise to fractional powers of $M$, while the triangle part contains
low-energy singularities.
The appearance of these graphs does not constitute a problem 
since their low-energy singularities are not new types of singularities but 
rather originate from a lower number of loops. Even after separating out $R_{G}$, 
all the thresholds and the associated analytic structure of these low-energy 
singularities are still contained in the low-energy part $\bar{I}_{G}$ through 
subgraphs with fewer loops. In the two-loop example shown in Fig.~\ref{fig:twoloop},
the low-energy singularities are identical to those of the one-loop 
triangle graph which are retained in $\bar{I}_{G}$ according to 
the regularization scheme and the contribution of  Fig.~\ref{fig:twoloop}
corresponds to the triangle graph with a renormalized  four-heavy-particle 
vertex shown in Fig.~\ref{fig:twoloop}(b).

\section{Summary}\label{summary}

In summary, a manifestly Lorentz-invariant regularization scheme which 
does not violate power counting was derived.  This scheme separates a
Feynman integral $I_{G}$ represented by a Feynman graph $G$ into 
two parts, $I_{G}=\bar{I}_{G}+R_{G}$, and is referred to as EFTDR; the
entire low-energy analytic structure of $I_{G}$ is collected into
$\bar{I}_{G}$ while all the low-energy power counting violating
contributions are collected into the regular part $R_{G}$.  This
separation is achieved by identifying the power counting violating
contributions to particular solutions of 
the Landau equations which are represented by subgraphs of $G$.  In
EFTDR, each subgraph $g$ of $G$ is associated with a 
domain of integration $\Delta_{g}$ in the space of Feynman parameters such that the
power counting violating terms are separated out by extending the
domain of integration.  Simple diagrammatic rules were given which
allow the construction of $\bar{I}_{G}$ (and $R_{G}$) for one-loop diagrams with an
arbitrary number of heavy propagating particles.  $R_{G}$ can be
accounted through a renormalization of the LECs of the EFT Lagrangian
and discarded; the EFTDR regularized Feynman graph $I_{G}$ is then
$\bar{I}_{G}$.  It was shown that $R_{G}$ and $\bar{I}_{G}$ transform
separately under the transformations that leave the EFT Lagrangian
invariant such that all the Ward identities are satisfied by the EFTDR 
regularized Greens functions of the EFT.  The application of EFTDR to higher loops
was discussed and it was emphasized that the power counting violating
contributions can still be separated out systematically by extending the
domain of integration in the space of Feynman parameters.  EFTDR was
applied to the one-heavy particle and two-heavy 
particle sectors to one loop and shown to reproduce the results in
Refs.~\cite{bl,glps}.

\section*{Acknowledgments}

The authors would like to thank J. L. Goity for his support
and J. D. Walecka for valuable comments on the manuscript.  They would
also like to thank Thomas Jefferson
National Accelerator Facility (Jefferson Lab) for their 
hospitality.  This work was supported by the National Science
Foundation through grant \#~HRD-9633750 (DL and GP) and by the Deutsche 
Forschungsgemeinschaft through project Le~1189/1-1 (DL).

\begin{appendix}
  
  \section{Minimally contracted regular subgraphs} \label{sec:MCRSG}
\begin{figure}[tbp]
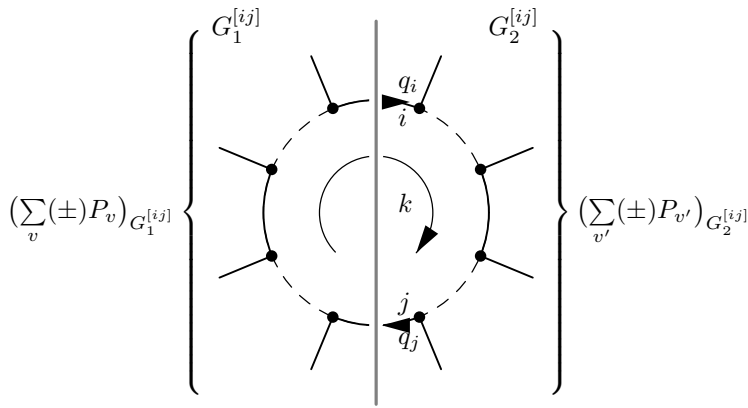

  \begin{center}
      \CutGraph
      \caption{Cutting a one-loop graph}
      \label{fig:mcrsg_cut}
    \end{center}
  \end{figure}

\begin{figure}[tbp]
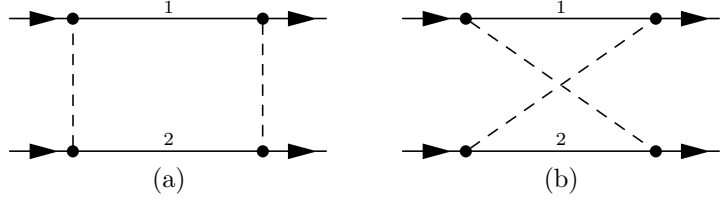

    \begin{center}
      \begin{tabular}{c@{\hspace{1cm}}c}
        \BoxGraph &\CrossBoxGraph \\[5ex]
        (a) & (b)
      \end{tabular}
      \caption{Box graph (a) and cross box graph (b)}
      \label{fig:BoxGraph}
    \end{center}
  \end{figure}

  Consider a general one-loop graph $G$ formed with heavy (mass $M$)
  and light (mass $m$) particles and external momenta $P_{v}^{\mu}$.
  By cutting two different lines $i$ and $j$ the graph is divided into two 
  connected parts $G_{1}^{[ij]}$ and $G_{2}^{[ij]}$, see
  Fig.~\ref{fig:mcrsg_cut}.
  Four-momentum conservation at each vertex of $G$ implies that the
  squared momentum flowing  through the cut, $(q_{i}-q_{j})^{2}$,
  equals the squared total external  momentum attached to either one
  of the two pieces,  
  \begin{equation}
    \label{eq:qiqj_def}
    (q_{i}-q_{j})^{2} = \bigl(\sum_{v} (\pm) P_{v}\bigr)_{G_{1}^{[ij]}}^{2} 
    =  \bigl(-\sum_{v'} (\pm) P_{v'}\bigr)_{G_{2}^{[ij]}}^{2}\:,
  \end{equation}
  where the momenta $q_{i}^{\mu}$ of internal lines are routed to flow 
  around the loop and the sign of the external momenta $P_{v}^{\mu}$
  depends on whether the momenta are incoming or outgoing.
  Since in an EFT all external three-momenta are ${\mathcal{O}}(p^2)\ll
  \Lambda_{{\text{EFT}}}$ and  creation/annihilation of heavy particles is
  excluded, it follows immediately that\footnote{%
    It is to be noted that Eq.~\eqref{eq:qiqj} remains true for all subgraphs
    of $G$. Indeed,  the l.h.s.\ is unaffected by the contraction lines to a point.} 
  \begin{equation}
    \label{eq:qiqj}
    (q_{i}-q_{j})^{2} = 
    \begin{cases}
      {\mathcal{O}}(p^{2}) & \text{$i$, $j$ light}\\
      M^{2} + {\mathcal{O}}(p) M &  \text{$i$ light, $j$  heavy}\qquad\qquad\\
      (2M)^{2} + {\mathcal{O}}(p) M & \text{$i$, $j$  heavy} \hfill\text{[a]}\\
      {\mathcal{O}}(p^{2}) & \text{$i$, $j$  heavy} \hfill\text{[b]}\\
    \end{cases}\:.
  \end{equation}
  An example for the case (\ref{eq:qiqj}a) is found when cutting the
  two heavy-particle lines of the box graph Fig.~\ref{fig:BoxGraph}(a), while
  the crossed box graph Fig.~\ref{fig:BoxGraph}(b) corresponds to case
  (\ref{eq:qiqj}b).

\begin{figure}[tbp]
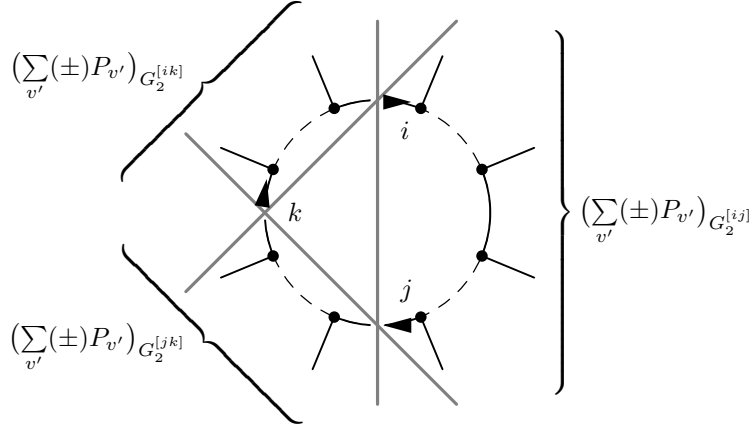

    \begin{center}
      \TransitivGraph
      \caption{Transitivity of cuts}
      \label{fig:mcrsg_transitiv}
    \end{center}
  \end{figure}

  \noindent
  The crucial observation is now that for any three heavy-particle internal
  lines $i$, $j$, and $k$:
  \begin{equation} \label{eq:mcrsg}
    \left.
      \begin{split}
        &(q_{i}-q_{j})^{2} = {\mathcal{O}}(p^2)\\
        &\text{and}\\
        &(q_{j}-q_{k})^{2} = {\mathcal{O}}(p^2)
      \end{split} \right\} \Rightarrow
    (q_{i}-q_{k})^{2} =  {\mathcal{O}}(p^2)\:.
  \end{equation} 
  This transitivity relation follows from Eq.~\eqref{eq:qiqj_def} and 
  \begin{equation}
    \label{eq:mcrsg2}
    \bigl(-\sum_{v} (\pm) P_{v}\bigr)_{G_{2}^{[ik]}} =  
    \bigl(\sum_{v} (\pm) P_{v}\bigr)_{G_{2}^{[ij]}} + 
    \bigl(\sum_{v} (\pm) P_{v}\bigr)_{G_{2}^{[jk]}}\:, 
  \end{equation}
  see Fig.~\ref{fig:mcrsg_transitiv}.
  Any two heavy-particle lines $i$ and $j$ of a graph $G$ are said to belong to
  the same \emph{minimally contracted regular} (MCR) subgraph ${\mathfrak{g}}$ if and only
  if they satisfy condition Eq.~(\ref{eq:qiqj}b),
  \begin{equation}
    \label{eq:mcrsg3}
    i,\:j \in {\mathfrak{g}} \quad :\Leftrightarrow (q_{i} - q_{j})^{2} = {\mathcal{O}}(p^{2})\:. 
  \end{equation}
 This defines the MCR subgraphs of $G$ completely.
 In light of Eq.~\eqref{eq:mcrsg}, different MCR subgraphs  cannot have any
 internal lines in common. On the other hand, every heavy-particle internal
 line belongs at least to one regular subgraph, namely the tadpole
 graph corresponding to that line. Therefore, the set of heavy-particle
 internal lines of a graph $G$ can be split into disjoint subsets;
 each subset is diagrammatically represented by a subgraph ${\mathfrak{g}}$ of $G$
 that consists of all the lines belonging to the subset and of no
 other lines. These subgraphs are the MCR subgraphs.
 They do not have low-energy singularities, since  by construction all
 (effective) external momenta are small while all internal lines are
 heavy.
 
 In the box graph Fig.~\ref{fig:BoxGraph}(a) the two heavy-particle lines 
 1 and 2 do \emph{not} belong to the same MCR subgraph; in this case
 the tadpole graphs formed with lines 1 and 2 are both MCR subgraphs. 
 In the cross box graph Fig.~\ref{fig:BoxGraph}(b), however, there is only \emph{one}
 MCR subgraph, namely the bubble graph formed with lines 1 and 2.

 It is noted that at most two different MCR subgraphs of a given
 one-loop graph exist. This can be seen as follows: assume the
 existence of more than two different MCR subgraphs and choose three
 heavy internal lines $i$, $j$, and $k$ belonging to different
 MCR subgraphs. It is easily verified that 
 \begin{equation} \label{eq:class2}
   \left.
     \begin{split}
       &(q_{i}-q_{j})^{2} = 4 M^{2} + {\mathcal{O}}(p) M\\
       &\text{and}\\
       &(q_{j}-q_{k})^{2} = 4 M^{2} + {\mathcal{O}}(p) M
     \end{split} \right\} \Rightarrow
   (q_{i}-q_{k})^{2} \neq 4 M^{2} + {\mathcal{O}}(p) M \:.
 \end{equation}
 However, since $(q_{i}-q_{k})^{2}$ must satisfy Eq.~\eqref{eq:qiqj},
 lines $i$ and $k$ have to belong to the same MCR subgraph,
 \begin{equation}
   (q_{i}-q_{k})^{2} = {\mathcal{O}}(p^2)\:,
 \end{equation}
 in contradiction to the above assumption.

\section{One-loop examples}\label{oneloopexamples}

\begin{figure}[htbp]
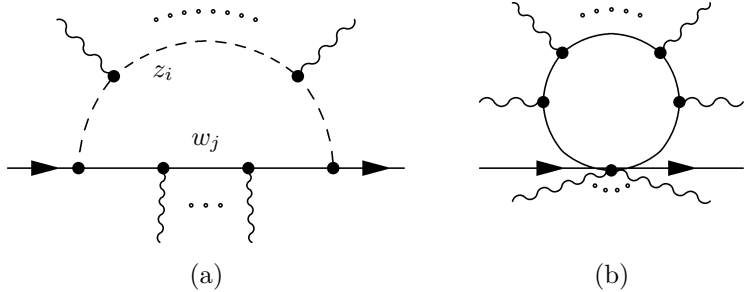

  \begin{center}
    \begin{tabular}{c@{\hspace{1cm}}c}
      \GenSelfenergyGraph &\RegularOneHeavyGraph \\
      (a) & (b) 
    \end{tabular}
    \caption{(a) General one-loop graph with one heavy particle. 
      (b) The corresponding minimally contracted regular subgraph.}
    \label{fig:GenSelfenergyGraph}
  \end{center}
\end{figure}

\begin{figure}[tbp]
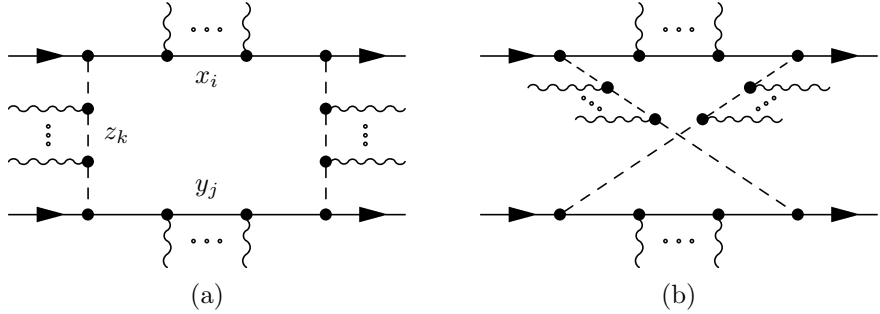

  \begin{center}
    \begin{tabular}{c@{\hspace{1cm}}c}
      \GenBoxGraph &\GenCrossBoxGraph\\
      (a) & (b)
    \end{tabular}
    \caption{ (a) Generalized box graph. (b) Generalized cross box graph.}
    \label{fig:TwoHeavy}
  \end{center}
\end{figure}

This formalism reproduces the prescriptions given for the one-loop
graphs in Figs.~\ref{fig:GenSelfenergyGraph}(a) and \ref{fig:TwoHeavy}(a)
first derived in Refs.~\cite{bl,glps}.
Fig.~\ref{fig:GenSelfenergyGraph}(a) is the most general one-loop
graph with a single heavy particle line running through it;
there are $H-1$ soft momentum insertions entering the heavy particle
internal line and $l-1$ soft momentum insertions entering the
light particle internal line.

The regular part for the graph in Fig.~\ref{fig:GenSelfenergyGraph}(a)
corresponds to the MCR
subgraph in Fig.~\ref{fig:GenSelfenergyGraph}(b) where all the light
internal lines have been contracted.
Eq.~\eqref{noncompactrep} gives
\begin{multline}
  \label{oneheavy}
  R=-\kappa_I\int_1^\infty\!\!\D \lambda\:
  \lambda^{H-1}(1- \lambda )^{l-1} \\ \times
  \int_0^1 \left(\prod\limits_{i=1}^l\mbox{d}z_i\right)
   \left(\prod\limits_{j=1}^{H}\mbox{d}w_j\right)  \delta(1-
  \sum\limits_{k=1}^lz_k)\delta(1-
  \sum\limits_{k=1}^{H}w_k) [C-\I\epsilon]^{d/2-I},
\end{multline}
which is the result first obtained by Becher and Leutwyler \cite{bl}
for a one-loop graph with $l$ light particle propagators labeled
$1,\dots,l$ and $H$ heavy particle propagators labeled $l+1,\dots,I$
where $I=l+H$.  Indeed, Eq.~\eqref{oneheavy}
describes a Feynman parameterization where all the light particle
propagators were combined into a single denominator, all the heavy
particle propagators were combined into another denominator, and these
two denominators were subsequently combined with the Feynman
parameter $\lambda$ integrated between one and infinity; this is
precisely the regular part in Becher-Leutwyler's infrared regularization.

For a Feynman graph with two heavy particle lines running through it,
Figs.~\ref{fig:TwoHeavy}(a)-(b) are the most general diagrams that can
be drawn.\footnote{For example, the triangle graph is obtained from
Fig.~\ref{fig:TwoHeavy}(a) by removing one of the light particle
internal lines that runs through the graph.}
For the Feynman graph shown in Fig.~\ref{fig:TwoHeavy}(a),
there are two MCR subgraphs obtained by contracting all the light particle
internal lines as well as one of the heavy
internal line running through the graph.  The regular part is given by 
the sum of the MCR subgraphs.
Assuming $l$ light particle internal lines and that the top heavy
particle internal line is segmented by $h_{1}-1$ soft momentum
insertions while the bottom heavy particle internal line is segmented
by $h_{2}-1$ soft momentum insertions, a change of variables similar
to Eq.~\eqref{cov} can be performed on Eq.~\eqref{noncompactrep}
to cast it in the form in which it appears in Ref.~\cite{glps}.
For the MCR obtained by contracting the $h_{n}$ ($n=1,2$) propagators
of one of the heavy lines, inserting the integrals 
\begin{align}
  1=\int_1^\infty\!\!\D \lambda \D \beta \: 
  \delta(\lambda-\sum\limits_{k=l+1}^I\alpha_k) \:
  \delta(\beta- \frac{1}{\lambda}\sum\limits_{k=l+h_{n}+1}^I\alpha_k) 
\end{align}
and performing the change of variables
\begin{alignat}{2}
  {\alpha}_{i} &= (1-\lambda)z_i &\qquad i&=1,\dots,l \nonumber \\
  {\alpha}_{i+l} &= \lambda(1-\beta)  x_i &\qquad i&=1,\dots,h_{n}\\
  {\alpha}_{i+l+h_{n}} &= \lambda\beta y_i &\qquad i&=1,\dots,H-h_{n}
  \nonumber 
\end{alignat}  
gives
\begin{multline}
  R=\sum\limits_{n=1,2}-\kappa_I\int_1^\infty\!\!\D \lambda\: 
\int_1^\infty\!\!\D \beta \:
\lambda^{H-1}(1- \lambda )^{l-1}\beta^{H-h_{n}-1}(1-\beta)^{h_{n}-1} \\ \times
  \int_0^1   \left(\prod\limits_{i=1}^{h_{n}}\mbox{d}x_i\right)
  \left(\prod\limits_{j=1}^{H-h_{n}}\mbox{d}y_j\right) 
  \left(\prod\limits_{k=1}^l\mbox{d}z_k\right)\\ \times
  \delta(1- \sum\limits_{a=1}^lz_a)
 \delta(1-  \sum\limits_{b=1}^{H-h_{n}}y_b) 
  \delta(1- \sum\limits_{c=1}^{h_{n}}x_c) 
  [C-\I\epsilon]^{d/2-I},
\end{multline}
where $H=h_{1}+h_{2}$.  This is just the regular part given in
\cite{glps}.  The generalized
cross box graph in Fig.~\ref{fig:TwoHeavy}(b) does not have a two-heavy
particle threshold, and its regular part is constructed from a single
MCR subgraph obtained by contracting all the light internal lines.

\end{appendix}

\end{fmffile}





\end{document}